\documentclass[conference,compsoc,final,twocolumn,10pt,a4paper]{IEEEtran}

\normalsize

\ifCLASSINFOpdf
\else
\fi

\usepackage[cmex10]{amsmath}
\usepackage{fixltx2e}
\usepackage{beamerarticle}
\renewcommand{\frametitle}[1]{}
\newcommand{\noteblock}[2]{}
\newcommand{\problemblock}[1]{}

\def\paper{gamma-lifting}

\usepackage{tabularx}

\usepackage[utf8]{inputenc}

\usepackage{xcolor}

\usepackage{csvsimple}

\usepackage{enumitem}
\SetEnumitemKey{hier}{label*=\arabic*., nosep, wide, leftmargin=*}

\usepackage{xfrac}
\usepackage{amsmath} 
\usepackage{amsfonts} 
\usepackage{amssymb}
\usepackage{trfsigns}

\usepackage{siunitx}
\DeclareSIUnit{\Baud}{Bd}
\DeclareSIUnit{\sym}{symbol}
\DeclareSIUnit{\belmilliwatt}{Bm}
\DeclareSIUnit{\dBm}{\deci\belmilliwatt}

\usepackage[nolist]{acronym} 
\begin{acronym}
 \acro{CE}{cross-entropy}
 \acro{SGD}{stochastic gradient descent}
 \acro{SCE}{static channel equalization}
 \acro{SC}{sub-carrier}
 \acro{NRZ}{non-return-to-zero}
 \acro{EEPN}{equalization enhanced phase noise}
 \acro{CT}{classical template}
 \acro{PMF}{probability mass function}
 \acro{RV}{random variable}
 \acro{CSI}{channel state information}
 \acro{UE}{user equipment}
 \acro{UL}{uplink}
 \acro{BS}{basestation}
 \acro{TDD}{time division duplex}
 \acro{FDD}{frequency division duplex}
 \acro{ECC}{error-correcting code}
 \acro{MLD}{maximum likelihood decoding}
 \acro{HDD}{hard decision decoding}
 \acro{IF}{intermediate frequency}
 \acro{RF}{radio frequency}
 \acro{SDD}{soft decision decoding}
 \acro{NND}{neural network decoding}
 \acro{CNN}{convolutional neural network}
 \acro{ML}{maximum likelihood}
 \acro{GPU}{graphical processing unit}
 \acro{BP}{belief propagation}
 \acro{LTE}{Long Term Evolution}
 \acro{BER}{bit error rate}
 \acro{SNR}{signal-to-noise-ratio}
 \acro{OSNR}{optical signal-to-noise-ratio}
 \acro{ReLU}{rectified linear unit}
 \acro{BPSK}{binary phase shift keying}
 \acro{QPSK}{quadrature phase shift keying}
 \acro{AWGN}{additive white Gaussian noise}
 \acro{MSE}{mean squared error}
 \acro{LLR}{log-likelihood ratio}
 \acro{MAP}{maximum a posteriori}
 \acro{NVE}{normalized validation error}
 \acro{BCE}{binary cross-entropy}
 \acro{BLER}{block error rate}
 \acro{SQR}{signal-to-quantisation-noise-ratio}
 \acro{MIMO}{multiple-input multiple-output}
 \acro{OFDM}{orthogonal frequency division multiplex}
 \acro{FDM}{frequency division multiplex}
 \acro{RF}{radio frequency}
 \acro{LOS}{line of sight}
 \acro{NLoS}{non-line of sight}
 \acro{NLI}{nonlinear interference}
 \acro{NMSE}{normalized mean squared error}
 \acro{CFO}{carrier frequency offset}
 \acro{SFO}{sampling frequency offset}
 \acro{IPS}{indoor positioning system}
 \acro{TRIPS}{time-reversal IPS}
 \acro{RSSI}{received signal strength indicator}
 \acro{MIMO}{multiple-input multiple-output}
 \acro{ENoB}{effective number of bits}
 \acro{AGC}{automated gain control}
 \acro{ADC}{analog to digital converter}
 \acro{ADCs}{analog to digital converters}
 \acro{FB}{front bandpass}
 \acro{FPGA}{field programmable gate array}
 \acro{JSDM}{Joint Spatial Division and Multiplexing}
 \acro{NN}{neural network}
 \acro{IF}{intermediate frequency}
 \acro{LoS}{line-of-sight}
 \acro{DSP}{digital signal processing}
 \acro{AFE}{analog front end}
 \acro{SQNR}{signal-to-quantisation-noise-ratio}
 \acro{SINR}{signal-to-interference-noise-ratio}
 \acro{ENoB}{effective number of bits}
 \acro{AGC}{automated gain control}
 \acro{PCB}{printed circuit board}
 \acro{EVM}{error vector mangnitude}
 \acro{CDF}{cumulative distribution function}
 \acro{MRC}{maximum ratio combining}
 \acro{MRP}{maximum ratio precoding}
 \acro{MRT}{maximum ratio transmission}
 \acro{DL}{deep-learning}
 \acro{SISO}{single-input single-output}
 \acro{SGD}{stochastic gradient descent}
 \acro{CP}{cyclic prefix}
 \acro{MISO}{Multiple Input Single Output}
 \acro{LMMSE}{linear minimum mean square error}
 \acro{ZF}{zero forcing}
 \acro{USRP}{universal software radio peripheral}
 \acro{RNN}{recurrent neural network}
 \acro{GRU}{gated recurrent unit}
 \acro{LSTM}{long short-term memory}
 \acro{NTM}{neural turing machine}
 \acro{DNC}{differentiable neural computer}
 \acro{TCN}{temporal convolutional network}
 \acro{FCL}{fully connected layer}
 \acro{MANN}{memory augmented neural network}
 \acro{RNN}{recurrent neural network}
 \acro{SE}{spectral efficiency}
 \acro{CD}{chromatic dispersion}
 \acro{FIR}{finite impulse response}
 \acro{TBP}{time-bandwidth-product}
 \acro{AE}{autoencoder}
 \acro{SSFM}{split-step Fourier Method}
 \acro{KNL}{Kerr-nonlinearity}
 \acro{QAM}{quadrature-amplitude-modulation}
 \acro{DNN}{dense neural network}
 \acro{PSD}{power spectral density}
 \acro{ASE}{amplified spontaneous emission}
 \acro{NLSE}{nonlinear Schr\"odinger equation}
 \acro{GNLSE}{Generalized nonlinear Schr\"odinger equation}
 \acro{SSFM}{split-step Fourier method}
 \acro{DA}{digital to analog}
 \acro{DAC}{digital to analog converter}
 \acro{LPF}{lowpass filter}
 \acro{TX}{transmitter}
 \acro{RX}{receiver}
 \acro{IQ}{in-phase quadrature}
 \acro{DC}{direct-current}
 \acro{MI}{mutual information}
 \acro{PMD}{polarization mode dispersion}
 \acro{PMF}{probability mass function}
 \acro{DBP}{digital back propagation}
 \acro{NFDM}{nonlinear frequency-domain multiplexing}
 \acro{AIR}{achievable information rate}
 \acro{ISI}{inter-symbol interference}
 \acro{NLC}{nonlinearity compensation}
 \acro{NFT}{nonlinear Fourier transform}
 \acro{PE}{post-equalization}
 \acro{PD}{predistortion}
 \acro{SPM}{self phase modulation}
 \acro{WDM}{wavelength division multiplex}
 \acro{DSCM}{digital sub-carrier multiplexing}
 \acro{PS}{probabilistic shaping}
 \acro{UWB}{ultra-wideband}
 \acro{GVD}{group velocity dispersion}
 \acro{CNLSE}{coupled nonlinear Schr\"odinger equation}
 \acro{EDFA}{erbium-doped fiber amplifier}
 \acro{FWM}{four-wave mixing}
 \acro{XPM}{cross-phase modulation}
 \acro{GN}{Gaussian noise}
\end{acronym}

\usepackage{caption}
\usepackage{subcaption}

\usepackage{float}

\usepackage{listings}
\makeatletter
\def\lst@makecaption{%
	\def\@captype{figure}%
	\@makecaption
}
\makeatother
\usepackage{svg}
\svgpath{fig/svg/}
\svgsetup{%
	 inkscapeversion=1,%
	 inkscapepath=fig/svg-generated/,%
}

\usepackage{comment}

\newcommand{\results}{data/}
\newcommand{\training}[1]{\results #1/training_#1.csv}
\newcommand{\evaluation}[2]{\results #1/evaluation_#2.csv}

\usepackage{tikz}

\usetikzlibrary{arrows}
\usetikzlibrary{arrows.meta}
\usetikzlibrary{spy}
\usetikzlibrary{chains}
\usetikzlibrary{positioning}
\usetikzlibrary{patterns}
\usetikzlibrary{calc}
\usetikzlibrary{backgrounds}
\usetikzlibrary{colorbrewer}
\usetikzlibrary{decorations.pathreplacing}
\usetikzlibrary{decorations.text}
\usetikzlibrary{external}
\tikzsetexternalprefix{fig/tikz-generated/}
\tikzset{external/figure name={}}

\tikzstyle{block} = [draw, thick, rectangle, minimum height = 3em, minimum width = 3em, inner sep = 5pt, fill=white]
\tikzstyle{colorblock}= [draw, thick, rectangle, minimum height = 3em, minimum width = 3em, inner sep = 5pt, fill=olive]
\tikzstyle{source} = [draw, thick, circle, minimum size = 4em, inner sep = 0pt, fill=white]
\tikzstyle{sink} = [draw, thick, circle, minimum size = 4em, inner sep = 0pt, fill=white]
\tikzstyle{op} = [draw, circle, fill=white]
\tikzstyle{input} = [draw, circle, minimum size = 2mm, inner sep = 0pt, fill=white]
\tikzstyle{bullet} = [draw, circle, minimum size = 2mm, inner sep = 0pt, fill]
\tikzstyle{output} = [coordinate]
\tikzstyle{none} = 	[]

\newcommand{\blockbox}[2]{\begin{minipage}{#1}\centering#2\end{minipage}}

\usepackage{pgfplots}
\usepackage{pgfplotstable}
\usepgfplotslibrary{groupplots}
\usepgfplotslibrary{external} 
\usepgfplotslibrary{colormaps}  
\usepgfplotslibrary{fillbetween}
\pgfplotsset{layers/standard/.define layer set={background,
		axis background,axis grid,axis ticks,axis lines,axis tick labels,pre
		main,main,
		axis descriptions,axis foreground
	}{
		grid style={/pgfplots/on layer=axis grid},
		tick style={/pgfplots/on layer=axis ticks},
		axis line style={/pgfplots/on layer=axis lines},
		label style={/pgfplots/on layer=axis descriptions},
		legend style={/pgfplots/on layer=axis descriptions},
		title style={/pgfplots/on layer=axis descriptions},
		colorbar style={/pgfplots/on layer=axis descriptions},
		ticklabel style={/pgfplots/on layer=axis tick labels},
		axis background@ style={/pgfplots/on layer=axis background},
		3d box foreground style={/pgfplots/on layer=axis foreground},
}}
\pgfplotsset{%
	curve/.append style={
		thick,  
		mark size=3pt, 
		mark repeat=1, 
		mark phase=0
	}
}

\pgfplotsset{%
evm chart/.append style={
	title={}, 
	width=1.0\columnwidth,
	height=20em,
	font=\rmfamily\small,
	legend style={font=\rmfamily\small, fill opacity=0.9, text opacity=1.0, cells={align=left}},
	legend columns=1,
	legend pos=south west,
	axis line style = thick,
	legend cell align={left},
	grid=major,
	xlabel near ticks,
	ylabel near ticks,
	xmin=-30.0,   xmax=10,
	ymin=10,   ymax=40,
	nodes near coords={
	},
}
}

\pgfplotsset{%
	se chart/.append style={
		title={}, 
		width=1.0\columnwidth,
		height=20em,
		font=\rmfamily\small,
		legend style={font=\rmfamily\small, fill opacity=0.9, text opacity=1.0, cells={align=left}},
		legend columns=1,
		legend pos=south west,
		axis line style = thick,
		legend cell align={left},
		grid=major,
		xlabel near ticks,
		ylabel near ticks,
		xmin=-30.0,   xmax=10,
		ymin=0,   ymax=9,
		nodes near coords={
		},
	}
}

\newcommand*{\flextitle}[4][]{%
	\ifdefined\report
		\chapter{#3}%
	\fi
	\ifdefined\slides
		\def\authors{#2}
		\lecture{#3}{#4}
		\part{#3}
	\fi
	\ifdefined\paper
		\title{#3\thanks{#1}}%
		\author{#2}%
		\maketitle%
	\fi
}

\tikzsetexternalprefix{fig/tikz-generated/article/\paper/}

\newcommand{\classiccdnl}{20210126113018}

\newcommand{\aecdnltwenty}{20200620154530}

\newcommand{\classiccdnlse}{\results\classiccdnl/evaluation_20210126113018.csv}

\newcommand{\aecdnltwentyse}{\results\aecdnltwenty/csv/-10dBm/se.csv}

\def\ps@IEEEtitlepagestyle{%
	\def\@oddfoot{\mysubmissionnotice}%
	\def\@evenfoot{}%
}
\def\mysubmissionnotice{%
	{\footnotesize  Accepted (09.03.2022) for presentation at the 23rd IEEE/ITG-Symposium on Photonic Networks, Berlin, Germany, 18-19.05.2022.\hfill}
	\gdef\mysubmissionnotice{}
}

\begin{document}
	
\flextitle{%
	\IEEEauthorblockN{%
		Tim Uhlemann, Alexander Span, Sebastian D\"orner, and Stephan ten Brink\\}%
	\IEEEauthorblockA{%
		\textit{Institute of Telecommunications}\\
		\textit{University of  Stuttgart}\\
		Pfaffenwaldring 47, 70569 Stuttgart, Germany\\
		\{uhlemann,span,doerner,tenbrink\}@inue.uni-stuttgart.de\\
	}}
	{Introducing $\gamma$-lifting for Learning Nonlinear\\Pulse Shaping in Coherent Optical Communication}
	{gamma-lifting}
\mode*


\begin{abstract}
Pulse shaping for coherent optical fiber communication has been an active area of research for the past decade. Most of the early schemes are based on classic Nyquist pulse shaping that was originally intended for linear channels.
The best known classic scheme, the \emph{split \acl{DBP}},
uses joint \acl{PD} and \acl{PE} and hence, a \emph{nonlinear} \acl{TX}; it, however, suffers from spectral broadening on the fiber due to the Kerr-effect.
With the advent of deep learning in communications, it has been realized that an ``\acl{AE}'' can learn to communicate efficiently over the optical fiber channel, jointly optimizing geometric constellations and pulse shaping -- while also taking into account linear and nonlinear impairments such as chromatic dispersion and Kerr-nonlinearity. E.g., \cite{uhlemann_deep-learning_2020} shows how an \acl{AE} can learn to mitigate spectral broadening due to the Kerr-effect using a trainable \emph{linear} \acl{TX}.
In this paper, we extend this \emph{linear} architectural template to a scalable \emph{nonlinear pulse shaping} consisting of a \acl{CNN} at both transmitter and receiver.
By introducing a novel $\gamma$-lifting training procedure tailored to the nonlinear optical fiber channel, we achieve stable \acl{AE} convergence to pulse shapes reaching information rates outperforming the classic split \acl{DBP} reference at high input powers.
\end{abstract}


\begin{figure*}[ht] 
\begin{subfigure}[c]{\textwidth}
	\centering
	\includegraphics[width=\textwidth]{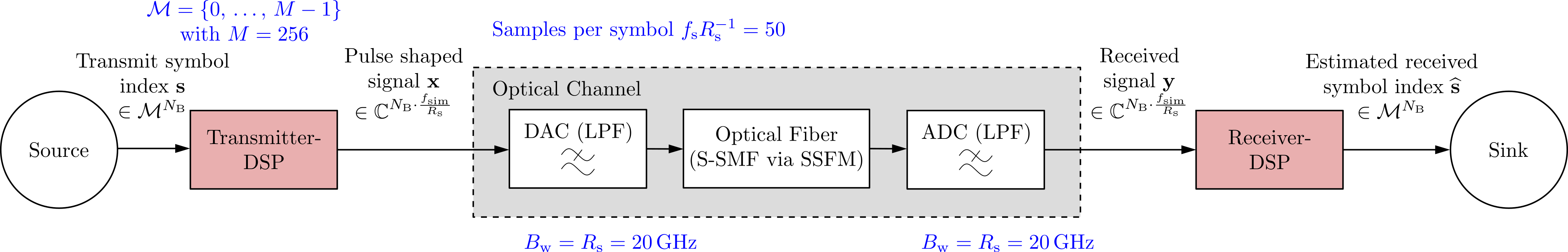}
	\subcaption{}
	\label{fig:system}
	\vspace{-.8em}
\end{subfigure}
\begin{subfigure}[c]{\textwidth}
	\centering
	\includegraphics[width=0.9\textwidth]{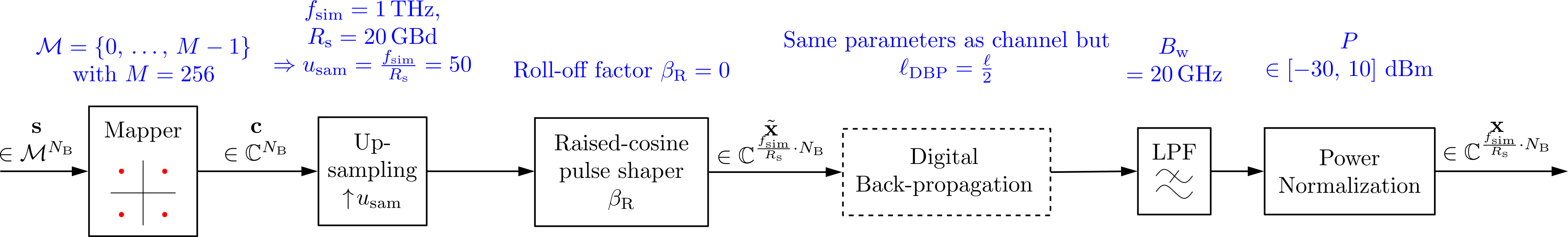}
	\subcaption{}
	\label{fig:dbp-tx}
	\vspace{-1.2em}
\end{subfigure}
\begin{subfigure}[c]{\textwidth}
	\centering
	\includegraphics[width=0.6\textwidth]{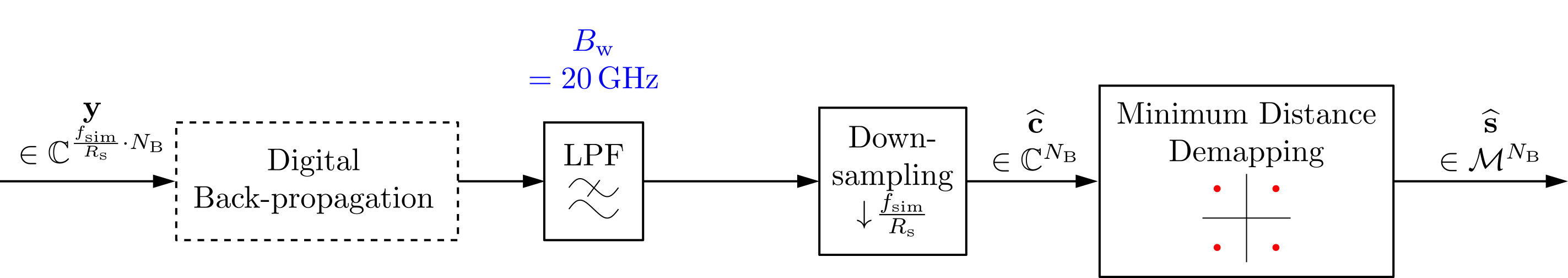}
	\subcaption{}
	\label{fig:dbp-rx}
\end{subfigure}
\caption{System model with (\subref{fig:system}) overview that holds for conventional as well as \acs{AE} setup, (\subref{fig:dbp-tx}) the \acs{DSP} for the conventional  \acs{TX}, and (\subref{fig:dbp-rx}) the \acs{RX} based on split \acs{DBP}.}
\end{figure*}

\mode<presentation>{

	\begin{frame}
		\frametitle{Agenda}
		\begin{itemize}
			\item The reference system
			\item A Nonlinear Architectural Template for the Autoencoder's Transmitter
			\item $\gamma$-lifted Training
			\item Conclusion and (as an Outlook) Extension to DWDM
		\end{itemize}
	\end{frame}
}

\mode<article>{\section{Introduction}}

For the design of an optical fiber communication system that achieves a high \ac{SE} one requires higher order modulation formats and thus a high \ac{SNR} at the \ac{RX}. This, in turn, needs higher launch powers at the \ac{TX}. Furthermore, a \ac{TBP} close to one should be chosen to not limit the \ac{SE}, i.e., symbol rate and occupied bandwidth should almost equal. This, in general, holds for the copper and wireless (radio frequency) channel. In contrast, for the optical fiber, a high input power results in a large Kerr-nonlinearity and hence, in a highly distorted signal at the \ac{RX} \cite{millar_mitigation_2010}. Apart from the signal-noise interaction, in a single channel scenario, this nonlinearity-induced distortion is deterministic and may be mitigated at \ac{TX} or \ac{RX}. Nevertheless, to obtain an optimal equalization performance, one is either forced to take into account the Kerr-nonlinearity-induced spectral broadening and, thus, support a wider bandwidth compared to the chosen symbol rate \cite{yousefi_per-sample_2011, xu_digital_2017}; or, operate in a regime where spectral broadening can be neglected \cite{kaminow_chapter_2002}. The first approach reduces \ac{SE}, whereas the second limits design and hardware options; For instance regarding the symbol rate, as spectral broadening inversely increases with it (as will be shown later).

Our superimposed goal is to develop a communication system that is able to overcome the need for additional bandwidth but still achieves optimal (or at least a better) performance than conventional reference systems. 

For those, \ac{DSP} algorithms have shown to outperform physical methods such as optical phase conjugation \cite{pepper_compensation_1980}, or optical back propagation \cite{bidaki_raman-pumped_2020}. Digital approaches can be further divided into three types, \emph{(i)} a static, analytical, and conventional equalization by fixed algorithms, such as the Volterra-series \cite{peddanarappagari_volterra_1997}, \ac{DBP} \cite{li_electronic_2008}, and perturbation models \cite{mecozzi_nonlinear_2012}; \emph{(ii)} \ac{NFT}-based schemes \cite{yousefi_information_2014-2}; or \emph{(iii)} \ac{DL} techniques with architectural templates based on the aforementioned conventional systems -- or based on, e.g., a \ac{NN} \cite{karanov_end--end_2018}.

As will be shown later, the \ac{DBP} is able to compensate for channel impairments almost perfectly up to a certain launch power where signal-noise interaction becomes relevant. This is already known and expected as it is the ideal inverse of the \ac{SSFM}-algorithm used to simulate the optical channel. Nevertheless, by limiting the bandwidth of \ac{TX} and \ac{RX} the performance decreases significantly due to spectral broadening only. This way we can show that the later introduced \ac{AE} is able to compensate for spectral broadening and, hence, allows to operate at higher input powers when compared to conventional systems. One particularity is that training the proposed structure requires a specific procedure that we refer to as ``$\gamma$-lifting'' to achieve stable convergence, and to find better local optima.

This paper, thereby, reports significant progress over \cite{uhlemann_deep-learning_2020}, where an \ac{AE} with linear \ac{TX} was trained to mitigate spectral broadening at high input powers. The training results there have shown that the \ac{AE} is still not fully able to compensate for the aforementioned nonlinear impairments at higher input powers $P>5\,\mathrm{dBm}$. 
Hence, here, we extend the \ac{AE} of \cite{uhlemann_deep-learning_2020} to include a \ac{NLC} architecture, effectively implementing a nonlinear \ac{TX}. Note that the linear \ac{TX} of \cite{uhlemann_deep-learning_2020} is still contained within the new model as a special case, allowing simple evaluation of the additional gains. The nonlinear extension is thereby inspired by the split \ac{DBP} consisting of a nonlinear and joint mitigation at \ac{TX} and \ac{RX}, currently providing the best performance \cite{lavery_benefit_2016}.

\ac{DL}-techniques are already widely used in almost all parts of an optical communication system. The novelty of this paper is the application of an architectural template with nonlinear \ac{TX}, i.e., its pulseshaping, as well as its particular training method.
In a similar spirit, e.g., Häger et al. \cite{hager_physics-based_2021}, proposed a trainable \ac{SSFM}, or \ac{DBP}, at the \ac{RX} of which the first half of the compensating path may be moved to the \ac{TX} and used as \ac{PD}. Despite the fact that our trainable architectural template is not based on the \ac{SSFM}, we train the \ac{PD} directly as part of the \ac{TX}. This allows the \ac{TX} to learn the opportunities of a \ac{PD} seeing all subsequent components. 
Neskorniuk et al. \cite{neskorniuk_end--end_2021} have used a trainable nonlinear \ac{PD} based on cubic correction terms over an approximated channel model applying the perturbation method. There, they have shown that the training can be significantly stabilized and accelerated up to a specific input power. However, we operate at input powers where this alternative channel model differs too much from the more accurate \ac{SSFM}, which we have applied in this work. Further, we also included the quasi-analog pulseshaping into the training.
Gaiarin et al. \cite{span_time-bandwidth_2019} have trained a transmitter based on the \ac{NFT} \cite{gaiarin_end--end_2021}, providing another nonlinear \ac{TX} structure. Yet, it has been shown that \ac{NFT}-based systems are currently far from spectrally efficient operations \cite{span_time-bandwidth_2019}.

The remainder of this work is structured as follows. Section~\ref{sec:system} provides a short description of the channel, the considered \ac{DBP}, as well as its performance drop at high launch powers.
In section~\ref{sec:autoencoder} the architectural template of the \ac{AE} is introduced including the nonlinear \ac{TX}. 
Section~\ref{sec:gamma-lifting} proposes a new training procedure tailored to the optical fiber channel that leads to improved equalization capabilities of the \ac{AE}. Finally, section~\ref{sec:summary} renders some conclusions and provides an outlook to future extensions.

\begin{table}[b] 
	\caption{Simulation parameters.}
	\begin{center}
		\begin{tabular}{c|c|c}
			\textbf{Property}&\textbf{Symbol}&\textbf{Value} \\
			\hline
			Planck's constant&$h$&$6.626\cdot 10^{-34}\,\si{\joule\second}$ \\
			Carrier frequency&$f_0$&$193.55\,\si{\tera\hertz}$ \\
			Attenuation&$\alpha$&$0.046\,\si{\per\kilo\meter}$ \\
			&&$\mathrel{\widehat{=}} 0.2\, \mathrm{dB}\,\mathrm{km}^{-1}$ \\
			Chromatic dispersion&$\beta_2$&$-21.67\,\si{\pico\second\squared\per\kilo\meter}$ \\
			Kerr-nonlinearity&$\gamma$&$1.27\,\si{\per\kilo\meter\per\watt}$ \\
			Spontaneous emission&$n_\mathrm{sp}$&$1$ \\
			Fiber length&$\ell$&$1000\,\si{\kilo\meter}$ \\
			Simulation sampling rate&$f_\mathrm{sim}$&$1\,\si{\tera\hertz}$ \\
			No. of SSFM-steps&$N_\mathrm{SSFM}$&$ 200 $ \\
			Fixed SSFM step-size&$\Delta z_\mathrm{SSFM}$&$ \SI{5}{\kilo\meter} $ \\
			Launch power&$P$&$-30\,\mathrm{dBm}, \,\ldots,\,10\,\mathrm{dBm}$ \\
			Bandwidth of TX/RX&$B_\mathrm{w}$&$20\,\si{\giga\hertz}$ \\
		\end{tabular}
		\label{tab:channel-parameters}
	\end{center}
\end{table}

\section{System model and performance references}\label{sec:system}

\mode<presentation>{
	\begin{frame}
		\frametitle{System: We assume a distributed Raman amplified S-SMF and single polarization.}
		\begin{figure}[ht]
			\centering
			\includesvg[width=\textwidth]{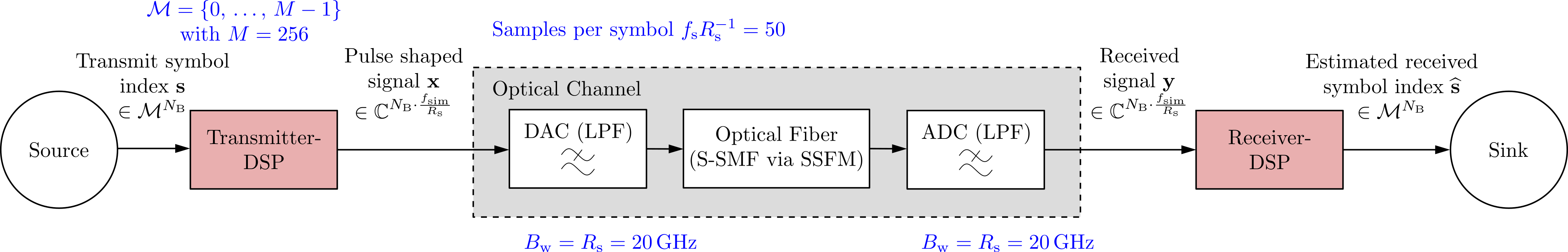}
			\label{fig:system-presentation}
			\vspace{-.8em}
		\end{figure}
		\small
		\begin{table}[b] 
			\begin{center}
				\begin{tabular}{c|c|c}
					\textbf{Property}&\textbf{Symbol}&\textbf{Value} \\
					\hline
					Chromatic dispersion&$\beta_2$&$\qty{-21.67}{\pico\second\squared\per\kilo\meter}$ \\
					Kerr-nonlinearity&$\gamma$&$\qty{1.27}{\per\kilo\meter\per\watt}$ \\
					Fiber length&$\ell$&$\qty{1000}{\kilo\meter}$ \\
					Simulation sampling rate&$f_\mathrm{sim}$&$\qty{1}{\tera\hertz}$ \\
					No. of SSFM-steps&$N_\mathrm{SSFM}$&$ 200 $ \\
					Fixed SSFM step-size&$\Delta z_\mathrm{SSFM}$&$ \qty{5}{\kilo\meter} $ \\
					Launch power&$P$&$\qty{-30}{\dBm}, \,\ldots,\,\qty{10}{\dBm}$ \\
					Bandwidth of TX/RX&$B_\mathrm{w}$&$20\,\si{\giga\hertz}$ \\
				\end{tabular}
				\label{tab:channel-parameters}
			\end{center}
		\end{table}	
	\end{frame}
	
	\begin{frame}
		\frametitle{Channel: Spectral broadening cannot be neglected here.}
		\centering
		\begin{table}
			\begin{tabular}{rl}
				Considered input power&$P=\qty{-30}{\dBm},\,\ldots,\,\qty{10}{\dBm}$ \\
				Dispersion length&$\ell_\mathrm{D}=\sfrac{T^2}{\left|\beta_2\right|} = \qty{115}{\kilo\meter}$ \\
				Nonlinear length&$\ell_\mathrm{NL}\left(P\right)=\sfrac{1}{\left(\gamma P\right)} = \qty{78e0}{\kilo\meter},\,\ldots,\,\qty{78e4}{\kilo\meter}$			
			\end{tabular}
		\end{table}
		Hence, $\ell_\mathrm{D}\geq\ell_\mathrm{NL}$ such that spectral broadening is not negligible \cite{essiambre_capacity_2010}, \cite{agrawal_nonlinear_2000}.
		\begin{figure}[ht]
			\centering
			\includesvg[width=0.8\textwidth]{length-comparison}
			\label{fig:length-comparison}
		\end{figure}
	\end{frame}
	
	\begin{frame}
		\frametitle{Transceiver: The best known conventional system performs a split DBP.}
		\vspace{1em}
		\begin{columns}[]
			\column{0.35\columnwidth}
			\column{0.3\columnwidth}
			\centering
			\textit{``\Ac{SPM} can be eliminated, within experimental error, up to where the transmitter bandwidth limits are reached''} \cite{roberts_electronic_2006}
			\column{0.35\columnwidth}
		\end{columns}
		\vspace{2em}
		\begin{figure*}[ht] 
			\begin{subfigure}[c]{\textwidth}
				\centering
				\includesvg[width=0.9\textwidth, inkscapearea=page]{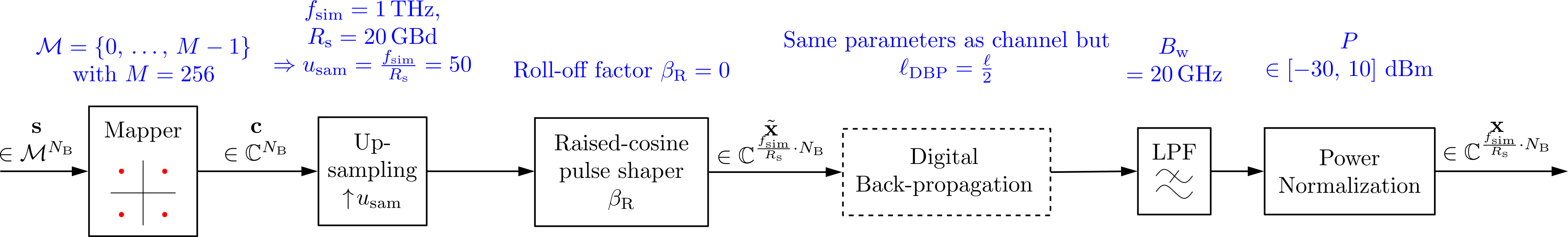}
				\label{fig:dbp-tx-presentation}
			\end{subfigure}
			\begin{subfigure}[c]{\textwidth}
				\centering
				\includesvg[width=0.6\textwidth, inkscapearea=page]{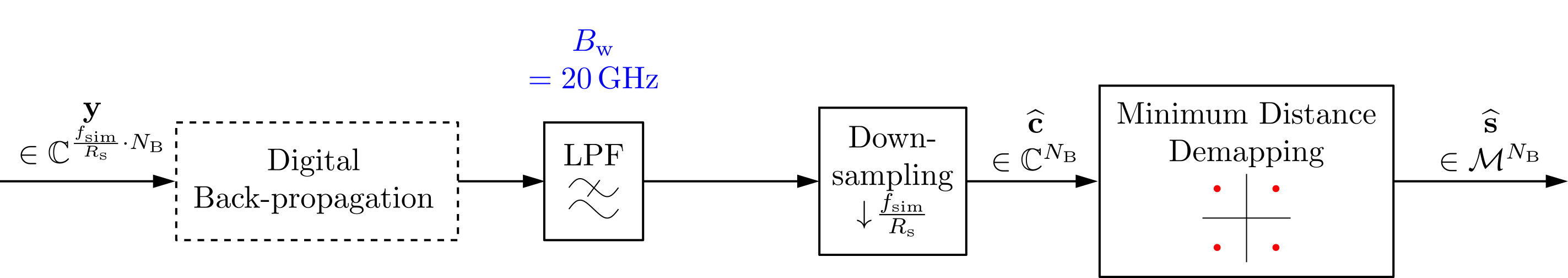}
				\label{fig:dbp-rx-presentation}
			\end{subfigure}
		\end{figure*}
	\end{frame}
	
	\begin{frame}
		\frametitle{Performance: Hard decision symbol-wise mutual information, or, spectral efficiency.}
		\textbf{Goal:} Include receiver's decisions into performance evaluation
		\begin{figure}
			\centering
			\includesvg[width=\textwidth, inkscapearea=page]{system-model-performance}
			\label{fig:system-model-performance-presentation}
		\end{figure}
	\end{frame}
}

The channel model used in this work is taken from \cite{uhlemann_deep-learning_2020} comprising the band-limited \ac{DAC} (represented by a \ac{LPF}), the optical fiber, and a band-limited \ac{ADC}, i.e., again an \ac{LPF} as shown in the dashed gray box of Fig.~\ref{fig:system}. All \acp{LPF} are ideally rectangular shaped. This model not only holds for the reference system but also for the \ac{AE} as introduced later.

The optical fiber is simulated via the symmetric \ac{SSFM} following the Wiener-Hammerstein model and, thus, approximating a solution of the \ac{NLSE} \cite{agrawal_nonlinear_2000}
\begin{equation}
\frac{\partial q(t,z)}{\partial z}= j\frac{\beta_2}{2}\frac{\partial^2 q(t,z)}{\partial t^2}-j\gamma |q(t,z)|^2 q(t,z) + n(t,z),
\label{eq:nlse}
\end{equation}
where ideal Raman amplification is assumed and the term accounting for fiber attenuation $\alpha$ was removed. Here $q(t,z)$ is the optical baseband signal with time $t$ and distance $z$. Alongside mitigating the attenuation, the amplification introduces noise $n(t,\,z)$ with power spectral density
\begin{equation}
\rho_\mathrm{n} = n_\mathrm{sp} h f_0 \alpha.
\end{equation}
The corresponding channel parameters are summarized in Tab.~\ref{tab:channel-parameters}. As we consider a single channel and single polarization system the only observable manifestation of the \mbox{Kerr-effect} is \ac{SPM}.

Among the fiber parameters, also some of the signal processing, e.g., those of the converters or \acp{DSP}, are fixed to obtain comparable systems; e.g., the supported bandwidth of both converters is set to $B_\mathrm{w}=20\,\mathrm{GHz}$. To avoid artificially limiting the \ac{SE} we have chosen a symbol rate of $R_\mathrm{s}=B_\mathrm{w}=20\,\mathrm{GBd}$ and hence, a symbol duration of $T=1/R_\mathrm{s}=50\,\mathrm{ps}$. The simulation bandwidth is $f_\mathrm{sim}=1\,\mathrm{THz}$ such that one obtains $f_\mathrm{sim}R_\mathrm{s}^{-1}=50$ samples per symbol. The resulting \ac{TBP} is 
\begin{equation}
	\Pi_\mathrm{TB}=T\cdot B_\mathrm{w}=1.
\end{equation}
The alphabet, or set of message symbol indices, is $\mathcal{M}=\left\{0,\,\ldots,\,M-1\right\}$ with $M=256$ and the number of transmitted symbols is $N_\mathrm{B}$.

As performance measure, we use the \acf{SE} defined as
\begin{equation}
\eta=\frac{I\left(s;\,\widehat{s}\right)}{T \cdot B_\mathrm{w}}=\frac{I\left(s;\,\widehat{s}\right)}{\Pi_\mathrm{TB}}\stackrel{\Pi_\mathrm{TB}=1}{=}I\left(s;\,\widehat{s}\right),
\end{equation}
where $I(s;\,\widehat{s})$ is the \ac{MI} between the distributions of the transmitted $s$ and the received (hard decided) symbol indices $\widehat{s}$ as shown in Fig.~\ref{fig:system}. Here we use histograms of hard decided symbols at \ac{RX} to include the whole \ac{RX}-\ac{DSP} in the chosen performance measure. Obviously, the performance could be improved by using soft outputs. But, in this work, we are more interested in studying how to mitigate the drop of \ac{MI} at high launch power and how to address this by means of pulse shaping.

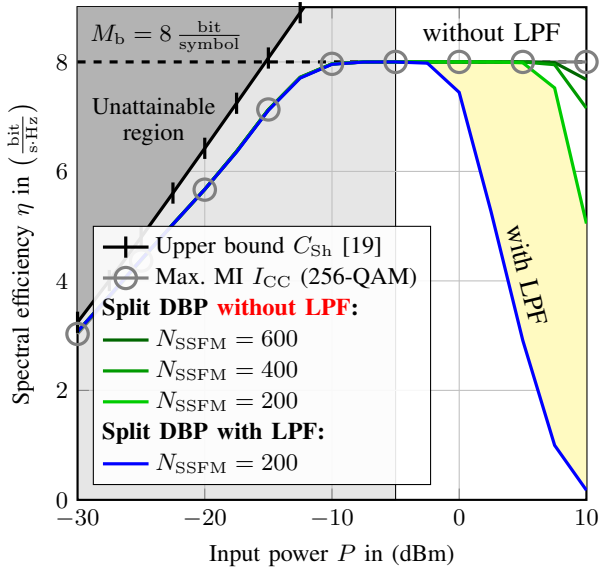
\begin{figure}[t] 
	\centering
	\tikzset{external/export next=false}
	\begin{tikzpicture}[scale=1.0, spy using outlines={rectangle, black, magnification=1.5, size=1.2cm, connect spies}]
		\begin{axis}[title={}, 
			width=1.0\columnwidth,
			height=23em,
			font=\small,
			legend cell align={left},
			legend style={font=\small, fill opacity=0.9, text opacity=1.0,},
			legend columns=1,
			legend pos=south west,
			axis line style = thick,
			xlabel={Input power $P$ in (dBm)},
			ylabel={Spectral efficiency $\eta$ in $\left(\frac{\mathrm{bit}}{\mathrm{s\cdot Hz}}\right)$},
			grid=major,
			xlabel near ticks,
			ylabel near ticks,
			xmin=-30.0,   xmax=10,
			ymin=0,   ymax=9,
			nodes near coords={
			},]
			
			\begin{pgfonlayer}{background}
				\draw[fill=black!10] (axis cs:-30, 0) rectangle (axis cs:-5, 9);
			\end{pgfonlayer}
			
			\addplot [very thick, forget plot, black, mark=none, dashed] coordinates {(-30, 8) (10, 8)};
			\node[above] at (axis cs:-23, 8) {\small{$M_\mathrm{b}=8\,\mathrm{\frac{bit}{symbol}}$}};
			
			\addplot[very thick, color=black, mark=|, mark size=4pt, mark repeat=1, mark phase=0, name path=SHANNON] table [x expr=10*log10(\thisrowno{33}*1000), y index = 326, col sep=semicolon] {\evaluation{20211119160424}{20211119160424}};
			\addlegendentry{{Upper bound $C_\mathrm{Sh}$ \cite{kramer_upper_2015}}};
			
			\addplot[draw=none, forget plot, mark=none, dashed, name path=INF] coordinates {(-30, 10) (10, 10)};
			\addplot[forget plot, black!30] fill between [of=INF and SHANNON];
			\node[above, align=center] at (axis cs:-24, 6.3) {Unattainable\\region};
			
			\addplot[very thick, color=gray, mark=o, mark size=4pt, mark repeat=2, mark phase=0] table [x expr=10*log10(\thisrowno{33}*1000), y index = 328, col sep=semicolon] {\evaluation{20211119160424}{20211119160424}};
			\addlegendentry{{Max. MI $I_\mathrm{CC}$ (256-QAM)}};
			
			\addplot[forget plot, draw=none, name path=A, no markers, mark options={decoration={name=none}}, decoration={text along path,
				text={{without LPF}},raise=+8pt,
				text align={left indent={0.74\dimexpr\pgfdecoratedpathlength\relax}},
			},
			postaction={decorate}] table [x expr=10*log10(\thisrowno{33}*1000), y index = 328, col sep=semicolon] {\evaluation{20211119160424}{20211119160424}};
			
			\addlegendimage{empty legend}
			\addlegendentry{\hspace{-.7cm}\textbf{Split DBP {\color{red}without LPF}:}};
			
			\addplot[very thick, color=green!40!black, mark=none, mark size=4pt, mark repeat=1, mark phase=0] table [x expr=10*log10(\thisrowno{33}*1000), y index = 327, col sep=semicolon] {\evaluation{20211217114913}{20211217114913}};			
			\addlegendentry{{$N_\mathrm{SSFM}=600$}};	
			
			\addplot[very thick, color=green!60!black, mark=none, mark size=4pt, mark repeat=1, mark phase=0] table [x expr=10*log10(\thisrowno{33}*1000), y index = 327, col sep=semicolon] {\evaluation{20211217103444}{20211217103444}};
			\addlegendentry{{$N_\mathrm{SSFM}=400$}};
			
			\addplot[very thick, color=green!80!black, mark=none, mark size=4pt, mark repeat=1, mark phase=0, name path=HIGH] table [x expr=10*log10(\thisrowno{33}*1000), y index = 327, col sep=semicolon] {\evaluation{20211217095804}{20211217095804}};
			\addlegendentry{{$N_\mathrm{SSFM}=200$}};
			
			
			\addlegendimage{empty legend}
			\addlegendentry{\hspace{-.7cm}\textbf{Split DBP with LPF:}};
			
			\addplot[very thick, color=blue, mark=none, mark size=4pt, mark repeat=2, mark phase=0, name path=LOW] table [x expr=10*log10(\thisrowno{33}*1000), y index = 328, col sep=semicolon] {\evaluation{20211119152924}{20211119152924}};
			\addlegendentry{{$N_\mathrm{SSFM}=200$}};
			
			\addplot[forget plot, yellow!30] fill between [of=HIGH and LOW];
			
			\addplot[forget plot, draw=none, name path=A, no markers, mark options={decoration={name=none}}, decoration={text along path,
				text={{with LPF}},raise=+4pt,
				text align={left indent={0.7\dimexpr\pgfdecoratedpathlength\relax}},
			},
			postaction={decorate}] table [x expr=10*log10(\thisrowno{33}*1000), y index = 328, col sep=semicolon] {\evaluation{20211119152924}{20211119152924}};
		\end{axis}
	\end{tikzpicture}
	\caption{Comparison of the performance of a split \acs{DBP} with and without (DAC/ADC-induced) \acsp{LPF}.}
	\label{fig:dbp-reference}
\end{figure}

\subsection{Spectral broadening revisited}

\mode<presentation>{
	\begin{frame}
		\frametitle{LPFs exhibit a limitation by bandwidth, or, spectral broadening, respectively.}
		\begin{columns}[onlytextwidth,T]
			\column{0.4\textwidth}
			\vspace{3em}
			\only<1->{			
				\begin{figure}
				\centering
				\includesvg[width=\textwidth]{system-model-no-lpfs}
				\label{fig:system-no-lpfs}
			\end{figure}
			}
			\only<2->{			
				\begin{figure}
					\centering
					\includesvg[width=\textwidth]{system-model-with-lpfs}
					\label{fig:system-with-lpfs}
				\end{figure}
			}
			\column{0.6\textwidth}
			\begin{figure}[t] 
				\centering
				\tikzset{external/export next=false}
				\begin{tikzpicture}[scale=1.0, spy using outlines={rectangle, black, magnification=1.5, size=1.2cm, connect spies}]
					\begin{axis}[title={}, 
						width=1.0\columnwidth,
						height=22em,
						font=\small,
						legend cell align={left},
						legend style={font=\small, fill opacity=0.9, text opacity=1.0,},
						legend columns=1,
						legend pos=south west,
						axis line style = thick,
						xlabel={Input power $P$ in (dBm)},
						ylabel={Spectral efficiency $\eta$ in $\left(\frac{\mathrm{bit}}{\mathrm{s\cdot Hz}}\right)$},
						grid=major,
						xlabel near ticks,
						ylabel near ticks,
						xmin=-30.0,   xmax=10,
						ymin=0,   ymax=9,
						nodes near coords={
						},]
						
						\only<1->{						
						\begin{pgfonlayer}{background}
							\draw[fill=black!10] (axis cs:-30, 0) rectangle (axis cs:-5, 9);
						\end{pgfonlayer}
						
						\addplot [very thick, forget plot, black, mark=none, dashed] coordinates {(-30, 8) (10, 8)};
						\node[above] at (axis cs:-23, 8) {\small{$M_\mathrm{b}=8\,\mathrm{\frac{bit}{symbol}}$}};
						
						\addplot[very thick, color=black, mark=|, mark size=4pt, mark repeat=1, mark phase=0, name path=SHANNON] table [x expr=10*log10(\thisrowno{33}*1000), y index = 326, col sep=semicolon] {\evaluation{20211119160424}{20211119160424}};
						\addlegendentry{{Upper bound $C_\mathrm{Sh}$ \cite{kramer_upper_2015}}};
						
						\addplot[draw=none, forget plot, mark=none, dashed, name path=INF] coordinates {(-30, 10) (10, 10)};
						\addplot[forget plot, black!30] fill between [of=INF and SHANNON];
						\node[above, align=center] at (axis cs:-24, 6.3) {Unattainable\\region};
						
						\addplot[very thick, color=gray, mark=o, mark size=4pt, mark repeat=2, mark phase=0] table [x expr=10*log10(\thisrowno{33}*1000), y index = 328, col sep=semicolon] {\evaluation{20211119160424}{20211119160424}};
						\addlegendentry{{Max. MI $I_\mathrm{CC}$ (256-QAM)}};
						
						\addplot[forget plot, draw=none, name path=A, no markers, mark options={decoration={name=none}}, decoration={text along path,
							text={{without LPF}},raise=+8pt,
							text align={left indent={0.74\dimexpr\pgfdecoratedpathlength\relax}},
						},
						postaction={decorate}] table [x expr=10*log10(\thisrowno{33}*1000), y index = 328, col sep=semicolon] {\evaluation{20211119160424}{20211119160424}};
						
						\addlegendimage{empty legend}
						\addlegendentry{\hspace{-.7cm}\textbf{Split DBP {\color{red}without LPF}:}};
						
						\addplot[very thick, color=green!40!black, mark=none, mark size=4pt, mark repeat=1, mark phase=0] table [x expr=10*log10(\thisrowno{33}*1000), y index = 327, col sep=semicolon] {\evaluation{20211217114913}{20211217114913}};			
						\addlegendentry{{$N_\mathrm{SSFM}=600$}};	
						
						\addplot[very thick, color=green!60!black, mark=none, mark size=4pt, mark repeat=1, mark phase=0] table [x expr=10*log10(\thisrowno{33}*1000), y index = 327, col sep=semicolon] {\evaluation{20211217103444}{20211217103444}};
						\addlegendentry{{$N_\mathrm{SSFM}=400$}};
						
						\addplot[very thick, color=green!80!black, mark=none, mark size=4pt, mark repeat=1, mark phase=0, name path=HIGH] table [x expr=10*log10(\thisrowno{33}*1000), y index = 327, col sep=semicolon] {\evaluation{20211217095804}{20211217095804}};
						\addlegendentry{{$N_\mathrm{SSFM}=200$}};}
						
						
						\only<2->{
						\addlegendimage{empty legend}
						\addlegendentry{\hspace{-.7cm}\textbf{Split DBP with LPF:}};
						
						\addplot[very thick, color=blue, mark=none, mark size=4pt, mark repeat=2, mark phase=0, name path=LOW] table [x expr=10*log10(\thisrowno{33}*1000), y index = 328, col sep=semicolon] {\evaluation{20211119152924}{20211119152924}};
						\addlegendentry{{$N_\mathrm{SSFM}=200$}};
						
						\addplot[forget plot, draw=none, name path=A, no markers, mark options={decoration={name=none}}, decoration={text along path,
							text={{with LPF}},raise=+4pt,
							text align={left indent={0.7\dimexpr\pgfdecoratedpathlength\relax}},
						},
						postaction={decorate}] table [x expr=10*log10(\thisrowno{33}*1000), y index = 328, col sep=semicolon] {\evaluation{20211119152924}{20211119152924}};
						
						\addplot[forget plot, yellow!30] fill between [of=HIGH and LOW];
					}
					\end{axis}
				\end{tikzpicture}
				\label{fig:dbp-reference}
			\end{figure}
	\end{columns}
	\end{frame}
}

Taking into account the considered power range for $P$, we obtain a dispersion length of
\begin{equation}
\ell_\mathrm{D}=T^2/|\beta_2|=115\,\mathrm{km}
\end{equation}
and a nonlinear length of
\begin{equation}
\ell_\mathrm{NL}(P)=1/(\gamma P)\in \left[78,\,78\cdot 10^{4}\right]\,\mathrm{km}
\end{equation}
such that $\ell_\mathrm{D} < \ell_\mathrm{NL}(P)$ does not hold for all input powers $P$ \cite{agrawal_nonlinear_2000}. This way we operate in a regime where spectral broadening cannot be neglected \cite{essiambre_capacity_2010}. Further it holds that
\begin{equation}
\ell_\mathrm{NL}(P)<\ell\ \forall\ P>-1.0\,\mathrm{dBm}.
\label{eq:nonlinear-length}
\end{equation}
Consequently, th received signal's bandwidth definitely exceeds the supported system bandwidth $B_\mathrm{w}=R_\mathrm{s}$ for conventional communication systems (and their corresponding equalization) as there is no margin implemented between $B_\mathrm{w}$ and the original signal bandwidth $R_\mathrm{s}$.\footnote{Note that a margin would artificially increase the \ac{TBP} and, hence, lower the \ac{SE}.} Next, we specify the goal for the \ac{AE} more precisely to ``Learning a high \ac{MI} despite the effect of spectral broadening.''

\begin{figure*}[ht] 
	\begin{subfigure}[c]{\textwidth}
		\centering
		\includegraphics[width=\textwidth]{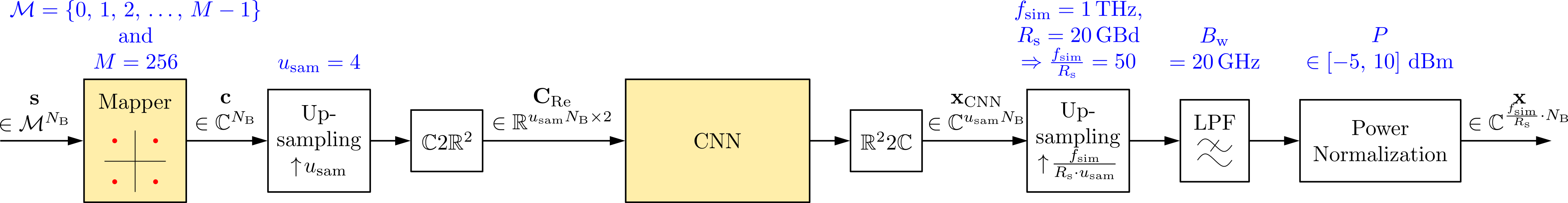}
		\subcaption{}
		\label{fig:aecnn-tx}
		\vspace{0.5em}
	\end{subfigure}
	\ifdefined\journal
	\begin{subfigure}[c]{\textwidth}
		\centering
		\includegraphics[width=0.7\textwidth]{fig/svg/aect-rx}
		\subcaption{}
		\label{fig:aecnn-rx-i}
	\end{subfigure}
	\fi
	\begin{subfigure}[c]{\textwidth}
		\centering
		\includegraphics[width=0.7\textwidth]{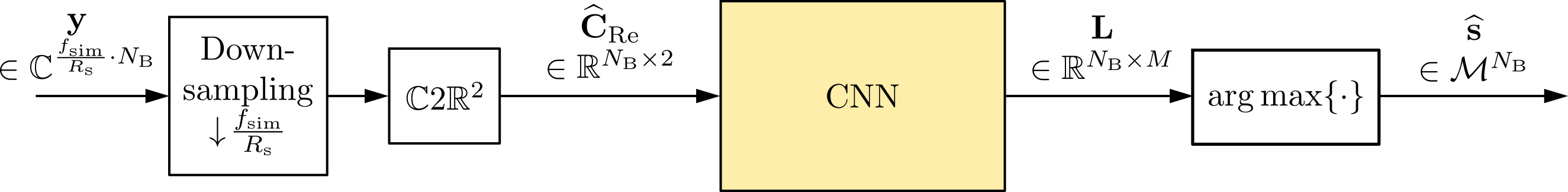}
		\subcaption{}
		\label{fig:aecnn-rx-ii}
	\end{subfigure}
	\caption{Architectural template of the proposed \acs{AE} implementing the \acs{TX}- and \acs{RX}-\acs{DSP}. Trainable blocks are colored in yellow. (\subref{fig:aecnn-tx}) shows the \acs{TX}-, and (\subref{fig:aecnn-rx-ii}) the \acs{RX}-\acs{DSP} as introduced in this work.}
	\label{fig:aecnn}
\end{figure*}

\subsection{Limitations of Digital Back Propagation}

As a reference system we use the ``split \ac{DBP}'' or also called ``split \ac{NLC}'' by \cite{lavery_benefit_2016}. The corresponding \ac{TX} and \ac{RX} are shown in Fig.~\ref{fig:dbp-tx} and Fig.~\ref{fig:dbp-rx}, respectively. 

The \ac{TX} consists of a conventional 256-\ac{QAM} that translates the symbol indices $\mathbf{s}\in\mathcal{M}^{N_\mathrm{B}}$ to complex constellation symbols $\mathbf{c}\in\mathbb{C}^{N_\mathrm{B}}$. An ideal Nyquist pulse-shaping (consisting of an upsampling to simulation frequency $f_\mathrm{sim}$ and a raised cosine pulse shaping with roll-off $\beta_\mathrm{R}=0$) lead to the bare transmit signal $\Tilde{\mathbf{x}}$. 
Typically a two-fold oversampling is used, whereas we use an oversampling of $u_\mathrm{sam}={f_\mathrm{sim}}/{R_\mathrm{s}}=50$ for a fair comparison with the later introduced trainable models that shall not learn to compensate any inaccuracy, and to exclude a numerical mismatch between channel and equalizer.
It follows an \ac{NLC} as \ac{PD} that is realized by a \ac{DBP}, which pre-compensates for a fiber length of $\ell_\mathrm{DBP}=\frac{\ell}{2}$. A subsequent \ac{LPF} plus power normalization generates the transmit signal $\mathbf{x}$.
The \ac{RX} starts with a \ac{DBP} processing the received signal $\mathbf{y}$ for a fiber length of again $\ell_\mathrm{DBP}=\frac{\ell}{2}$ at simulation rate, which is already lowpass-filtered by the channel (see converters in Fig.~\ref{fig:system}).\footnote{In the simulation the \ac{ADC} belongs to the channel, but technically the channel has no lowpass-characteristic.}An \ac{LPF} assures a correct downsampling to symbol rate $R_\mathrm{s}$ for a subsequent minimum distance demapping.
The \ac{DBP} algorithm itself is based on the symmetric \ac{SSFM} following the Wiener-Hammerstein model as shown in \cite{millar_mitigation_2010} that is also used for the channel simulation.

It was already shown that for a \ac{PD}-only implementation ``\ac{SPM} can be eliminated, within experimental error, up to where the transmitter bandwidth limits are reached'' \cite{roberts_electronic_2006}. Hence, in general the chosen \ac{PD} leads to a transmit signal that perfectly pre-compensates for channel impairments but occupies more bandwidth than the supported one by the converters. The same holds for the application of a \ac{PE}-only where now the receiver bandwidth limits the resulting performance. Hence, to overcome spectral broadening introduced by Kerr-nonlinearity, one has to take into account an artificial bandwidth expansion. This required expansion combined with the \ac{LPF} explains the performance drop of the split \ac{DBP} at high launch powers.

The curves of Fig.~\ref{fig:dbp-reference} confirms the statement above. In green it shows the \ac{SE} of the split \ac{DBP} without the aforementioned \acp{LPF}. Indeed, it achieves the expected near optimal performance up to a certain launch power, where the drop can be related to, either, the limited accuracy of the \ac{SSFM}, i.e., the number of steps $N_\mathrm{SSFM}$ used for the channel and the \ac{DBP}, or, signal-noise-interaction. To get a better understanding of the accuracy, this system is evaluated at different number of \ac{SSFM} steps. Enabling the \acp{LPF} one obtains the blue curve exhibiting a drop at lower launch powers than before. The space in between green and blue curve depicts the loss introduced only by spectral broadening.

For comparison \emph{(i)} an upper bound proposed by Kramer et al. \cite{kramer_upper_2015} that is equivalent to the Shannon limit and \emph{(ii)} the constellation constrained \ac{MI} $I_\mathrm{CC}$ for an \ac{AWGN} channel based on hard decisions and assuming a 256-\ac{QAM} are shown. The latter curve only indicates that, for low input powers, the conventional system achieves an almost optimal equalization. For high input powers, $I_\mathrm{CC}$ saturates as expected due to the constrained constellation.

\subsection{Discussion of Conventional Reference Curves}

Even if the \ac{DBP} currently shows the best performance, and indeed compensates for the nonlinear impairments almost perfectly, this only holds for the case where, either, spectral broadening is not relevant, or, \ac{DAC}, and/or \ac{ADC} support a larger bandwidth than required by the actual information, at the expense of a higher \ac{TBP} and, hence, lower \ac{SE} by design. 
This gives rise to the assumption that better solutions than \ac{DBP} in terms of \ac{SE} exist for the case where both converters significantly limit the bandwidth. I.e., the goal is to enter the attainable region between the curves with and without \ac{LPF}, indicated in yellow in Fig.~\ref{fig:dbp-reference}. While this goal has not changed since \cite{uhlemann_deep-learning_2020}, we now can apply both an extended AE and a new training procedure to further close this gap.

\mode<all>
\mode*
\section{Autoencoder-optimized pulse shaping}\label{sec:autoencoder}

\mode<presentation>{
	\begin{frame}
		\frametitle{The approach is to use an Autoencoder that learns to circumvent spectral broadening.}
		\begin{figure}[t]
			\centering
			\includesvg[width=\textwidth, inkscapearea=page]{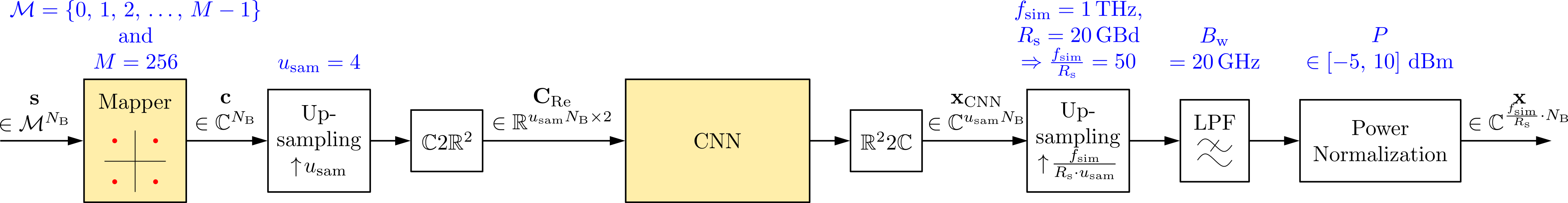}
			\label{fig:aecnn-tx-presentation}
		\end{figure}
		\only<2>{\begin{columns}[onlytextwidth, T]
			\column{0.5\textwidth}
			\begin{figure}[t]
				\includesvg[width=0.5\textwidth, inkscapearea=page]{classic}
				\label{fig:aecnn-tx-presentation}
			\end{figure}
		\vspace{1em}
		The CNN-structure is chosen based on the trainable linear filter from \cite{uhlemann_deep-learning_2020}.
			\column{0.5\textwidth}
			\begin{figure}
				\includegraphics[width=\textwidth]{fig/png/11}
			\end{figure}
		\end{columns}}
	\only<3>{\begin{columns}[onlytextwidth, T]
	\column{0.5\textwidth}
	\begin{figure}[t]
		\includesvg[width=\textwidth, inkscapearea=page]{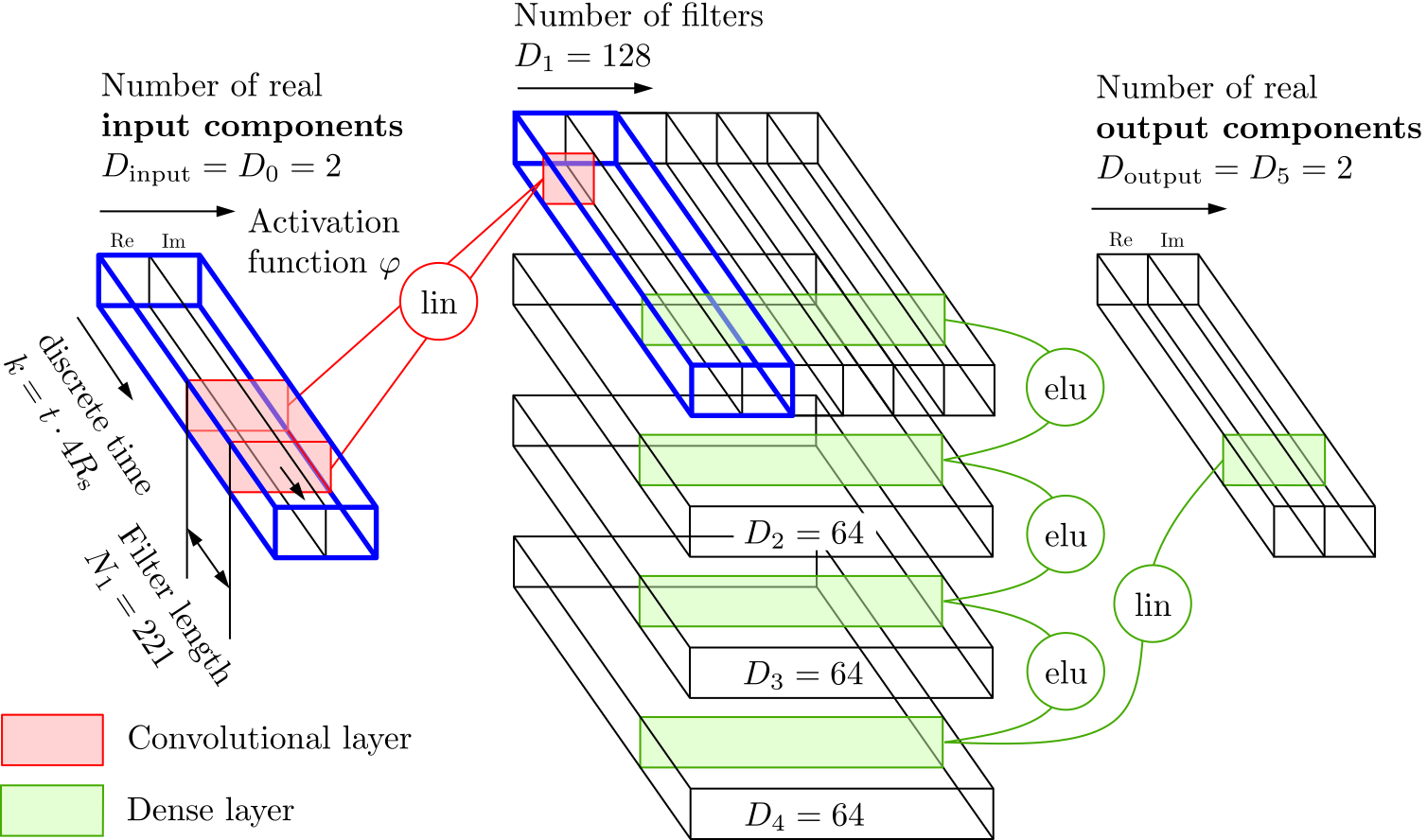}
		\label{fig:aecnn-tx-presentation}
	\end{figure}
	\column{0.5\textwidth}
	\begin{table}[b] 
		\small
		\begin{center}
			\begin{tabular}{c|c|c|c|c}
				\textbf{Layer}&\textbf{Type}&\multicolumn{3}{l}{\textbf{Parameters}} \\
				$\vartheta$&&$D_\vartheta$&$N_\vartheta$&$\varphi_\vartheta$ \\
				\hline
				1&Convolutional&128&221&linear\\
				2&Dense&64&-&elu \\
				3&Dense&64&-&elu \\
				4&Dense&64&-&elu \\
				5&Dense&2&-&linear \\
			\end{tabular}
		\end{center}
		\label{tab:cnn-configuration-presentation}
	\end{table}
\end{columns}}
	\end{frame}

	\begin{frame}
		\frametitle{The receiver uses the same structure based on CNNs without oversampling.}
		\begin{figure}[t]
			\centering
			\includesvg[width=0.7\textwidth, inkscapearea=page]{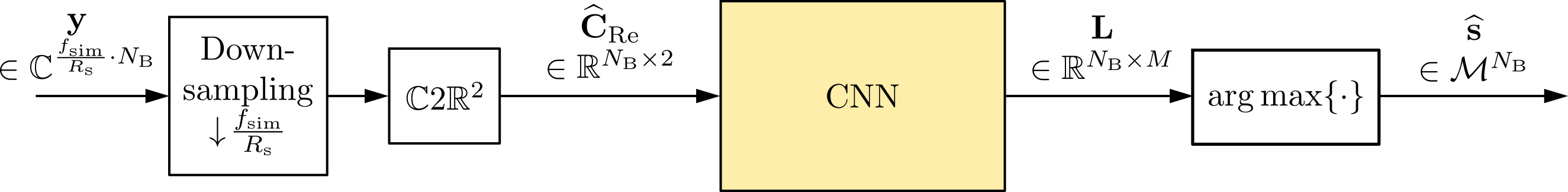}
			\label{fig:aecnn-rx-presentation}
		\end{figure}
\only<2>{	\begin{columns}[onlytextwidth, T]
	\column{0.5\textwidth}
	\begin{figure}[t]
		\includesvg[width=\textwidth, inkscapearea=page]{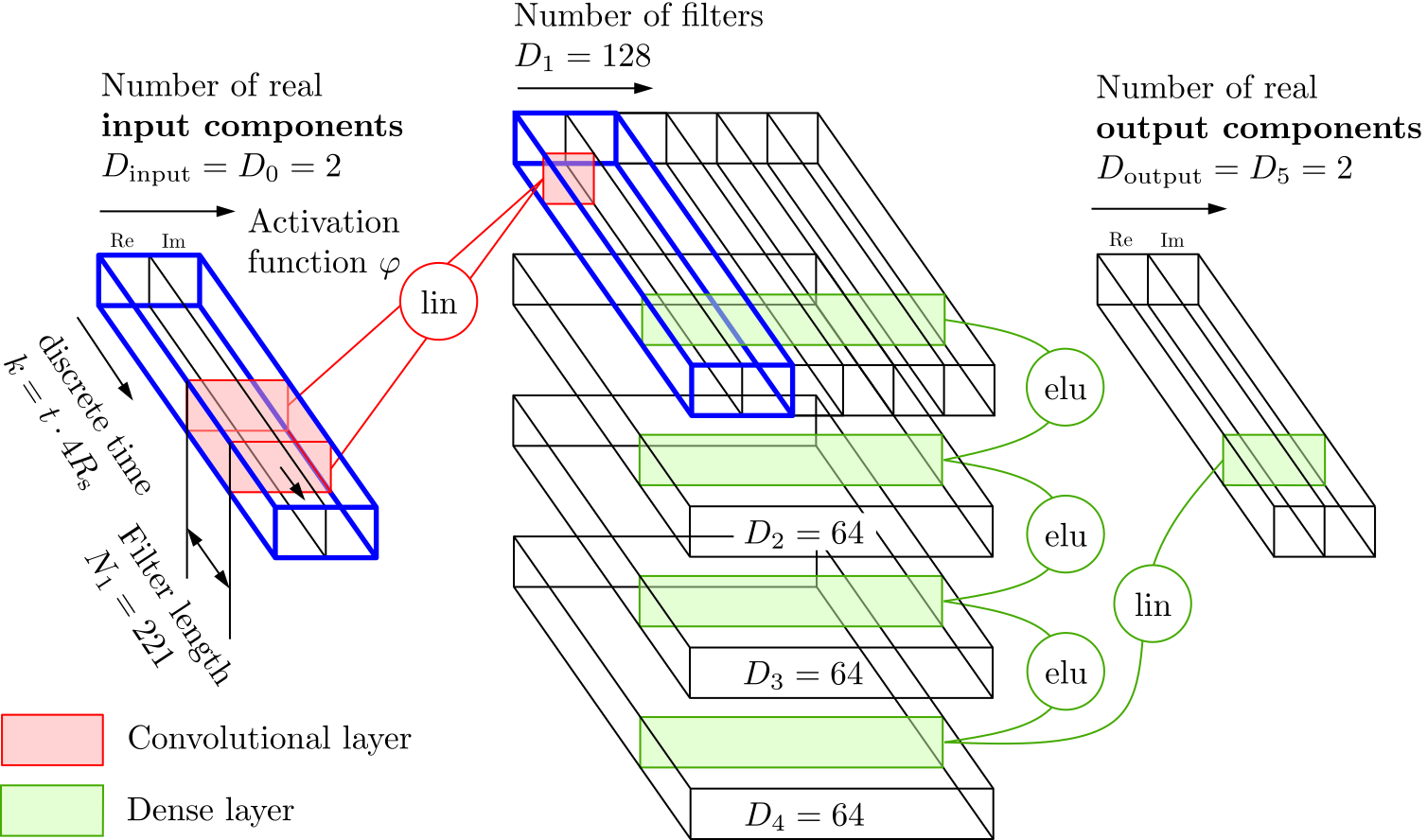}
		\label{fig:cnn-rx-presentation}
	\end{figure}
	\column{0.5\textwidth}
	\begin{table}[b] 
		\small
		\begin{center}
			\begin{tabular}{c|c|c|c|c}
				\textbf{Layer}&\textbf{Type}&\multicolumn{3}{l}{\textbf{Parameters}} \\
				$\vartheta$&&$D_\vartheta$&$N_\vartheta$&$\varphi_\vartheta$ \\
				\hline
				1&Convolutional&128&55&linear\\
				2&Dense&2048&-&elu \\
				3&Dense&2048&-&elu \\
				4&Dense&512&-&elu \\
				5&Dense&512&-&elu \\
				6&Dense&512&-&elu \\
				7&Dense&256&-&linear \\
			\end{tabular}
		\end{center}
		\label{tab:cnn-configuration-rx-presentation}
	\end{table}
\end{columns}}
	\end{frame}

	\begin{frame}
	\frametitle{For the end-2-end training we used cross-entropy loss on the classified symbols.}
	\begin{figure}[t]
		\includesvg[width=\textwidth, inkscapearea=page]{system-model-training}
		\label{fig:cnn-rx-presentation}
	\end{figure}
	\end{frame}
}

Fig.~\ref{fig:aecnn} shows the block diagram of the new architectural template for the \ac{AE} consisting of the \ac{TX} in Fig.~\ref{fig:aecnn-tx} and \ac{RX} in Fig.~\ref{fig:aecnn-rx-ii}, where now both are trainable and nonlinear.

\subsection{Brief recap of neural network notation}

A network consists of an ordered set of layers $\mathcal{L}=\left\{L_1,\,L_2,\,\ldots,\,L_\Theta\right\}$, where $L_\vartheta$ is the $\vartheta$th layer of the total number of $\Theta$ layers. Here, a layer is
\begin{equation}
	\begin{aligned}
		L_{\vartheta}: \mathbb{R}^{K_{\vartheta-1}\times D_{\vartheta-1}}&\rightarrow\mathbb{R}^{K_{\vartheta}\times D_{\vartheta}} \\
		\mathbf{U}_{\vartheta-1}&\mapsto \mathbf{U}_{\vartheta} \\
	\end{aligned}
\end{equation}
where $\mathbf{U}_{\vartheta-1}$ is the input and $\mathbf{U}_{\vartheta}$ is the output signal of the $\vartheta$th layer; $K_{\vartheta}\in\mathbb{N}$ is the length of the signal in terms of time samples after the $\vartheta$th layer; and $D_{\vartheta}$ is the dimensionality of each time sample, i.e., the number of channel components. An example of channel components are real and imaginary part of a signal with $D_\vartheta=2$. In the following $u_{\vartheta, k,d}=\left[\mathbf{U}_\vartheta\right]_{k,d}$ is element of a signal matrix at time instance $k$ and component $d$. In this work a layer is either a convolutional or a dense layer, where $\vartheta=0$ denotes the input signal.

For a convolutional layer $L_{\vartheta}=L_{\vartheta,\mathrm{conv}}$, the output dimension $D_\vartheta$ corresponds to the number of different trainable kernels or filters $\mathbf{H}_{\vartheta,d}\in\mathbb{R}^{N_\vartheta\times D_{\vartheta-1}}$ for $d=0,\,\ldots,\,D_\vartheta-1$; $N_\vartheta<K_\vartheta$ is the filter's individual length. All filters of a single layer have the same length. The output sample of a convolutional layer $\vartheta$ at time instance $k$ and for output component $d$ can be formulated as
\begin{equation}
	\begin{aligned}
		u_{\vartheta,k,d}=&L_{\vartheta,\mathrm{conv}}(\mathbf{U}_{\vartheta-1})\\
		=&\varphi_\vartheta\left\{\mathrm{tr}\left(\bar{\mathbf{U}}_{\vartheta-1,k}\cdot \mathbf{H}^\mathsf{T}_{\vartheta,d}\right)+b_{\vartheta,d}\right\} \\ 
	\end{aligned}
\label{eq:conv}
\end{equation}
where $\varphi_\vartheta$ is the activation function, $(\cdot)^\mathsf{T}$ is the transpose, $b_{\vartheta,d}$ is a trainable bias, and
\begin{equation}
	\begin{gathered}
		\bar{\mathbf{U}}_{\vartheta-1,k}=\left(\begin{array}{ccc} 
			u_{\vartheta-1,k-\bar{N},0} & \cdots & u_{\vartheta-1,k-\bar{N},D_{\vartheta-1}-1} \\ 
			\vdots & \ddots & \vdots \\ 
			u_{\vartheta-1,k+\bar{N},0} & \cdots & u_{\vartheta-1,k+\bar{N},D_{\vartheta-1}-1} \\ 
		\end{array}\right)\\
	\end{gathered}
	\label{eq:signal-slice}
\end{equation}
is a sliced and shifted version of the input signal with $\bar{N}=(N_{\vartheta}-1)/2$. The filter length $N_{\vartheta}$ is always chosen to be odd, to have a symmetric influence of the neighboring samples. A zero padding is added such that one obtains equal length of the input and output signal what is often referred to as ``same'' padding. This corresponds to $u_{\vartheta,k,d}=0\,\forall\, k<0\text{ or }k\geq K_{\vartheta}$.

For a dense layer $L_{\vartheta}=L_{\vartheta,\mathrm{dense}}$, the output dimension $D_\vartheta$ corresponds to the number of neurons of that layer and $N_\vartheta=1\,\forall\, \vartheta$ as, here, each layer operates on a single time instance $k$. This simplifies Eq.~(\ref{eq:conv}) to
\begin{equation}
	\begin{gathered}
		\begin{aligned}
			u_{\vartheta,k,d}&=L_{\vartheta,\mathrm{dense}}\left(\mathbf{U}_{\vartheta-1}\right) \\
			&=\varphi_\vartheta\left\{\left[\mathbf{U}_{\vartheta-1}\right]_k\cdot \mathbf{H}^\mathsf{T}_{\vartheta,d}+b_{\vartheta,d}\right\}.
		\end{aligned} \\
	\end{gathered}
\end{equation}

\subsection{Transmitter Design}

\begin{table}[b] 
	\caption{Parameters of the \acs{TX}- and \acs{RX}-\acs{NN}.}
	\small
	\begin{center}
		\begin{tabular}{c|c|c|c|c}
			\textbf{Layer}&\textbf{Type}&\multicolumn{3}{l}{\textbf{Parameters}} \\
			$\vartheta$&&$D_\vartheta$&$N_\vartheta$&$\varphi_\vartheta$ \\
			\hline
			\multicolumn{5}{l}{Transmitter}\\
			\hline
			1&Convolutional&128&221&linear\\
			2&Dense&64&-&elu \\
			3&Dense&64&-&elu \\
			4&Dense&64&-&elu \\
			5&Dense&2&-&linear \\
			\hline
			\multicolumn{5}{l}{Receiver}\\
			\hline
			1&Convolutional&128&55&linear\\
			2&Dense&2048&-&elu \\
			3&Dense&2048&-&elu \\
			4&Dense&512&-&elu \\
			5&Dense&512&-&elu \\
			6&Dense&512&-&elu \\
			7&Dense&256&-&linear \\
		\end{tabular}
	\end{center}
\label{tab:cnn-configuration}
\end{table}

Equivalent to the reference system, the symbol indices $\mathbf{s} \in \mathcal{M}^{N_\mathrm{B}}$ shall be transmitted by the \ac{TX} and estimated by the \ac{RX} with a symbol rate of $R_\mathrm{s}=20\,\mathrm{GBd}$. Here, the same system parameters hold as already given in Sec.~\ref{sec:system}. Each transmission contains a batch of $N_\mathrm{B}=12,000$ symbol indices, which is chosen on the basis of available GPU-memory in our simulation setup.

The \ac{TX} as given in Fig.~\ref{fig:aecnn-tx} starts with a trainable mapping, i.e., an $M \times 2$-matrix, choosing symbols $\mathbf{c}$ from a complex constellation based on the incoming symbol indices $\mathbf{s}$. Then, the signal is upsampled from symbol rate $R_\mathrm{s}$ to $u_\mathrm{sam}\cdot R_\mathrm{s}$ by adding zeroes in between the \ac{IQ}-symbols, and, subsequently, pulse shaped. Here, $u_\mathrm{sam}=4$ has shown to be sufficiently large to achieve the results in the following figures, staying well below the upsampling of the \ac{DBP}, where $u_\mathrm{sam}=50$.

The nonlinear pulse shaping generates the waveform $\mathbf{x}_\mathrm{CNN}$ and is realized by a \ac{CNN}. As such, it can only process real valued signals, and, thus, the two transformations $\mathbb{C}2\mathbb{R}^2$ and $\mathbb{R}^2 2\mathbb{C}$ vice-versa are required. Thereby, real and imaginary part of the complex signal are stacked in an additional last dimension. 
\ifdefined\journal
It was shown that native complex trainable structures can be implemented as well, exhibiting accelerated training convergence \cite{freire_complex-valued_2021}. Nevertheless, as the degree of freedom stays the same, we decided to stick to real implementations only.
\fi

It follows an upsampling from sample rate $u_\mathrm{sam}\cdot R_\mathrm{s}=80\,\mathrm{GHz}$ of the \ac{CNN} to $f_\mathrm{sim}=1\,\mathrm{THz}$, now including a Whittaker-Shannon interpolation for quasi-continuous waveform channel simulation. The lowpass-filtered and power normalized signal $\mathbf{x}$ is finally transmitted over the channel, with samples being calculated as $[\mathbf{x}]_k=g_\mathrm{TX}(k,\mathbf{s};\Omega_\mathrm{TX})$. Here, $g_\mathrm{TX}$ summarizes the \ac{TX} function with trainable parameters $\Omega_\mathrm{TX}$, i.e., the weights $\mathbf{H}_\vartheta$.

\begin{figure}[t] 
	\centering
	\includegraphics[width=\columnwidth]{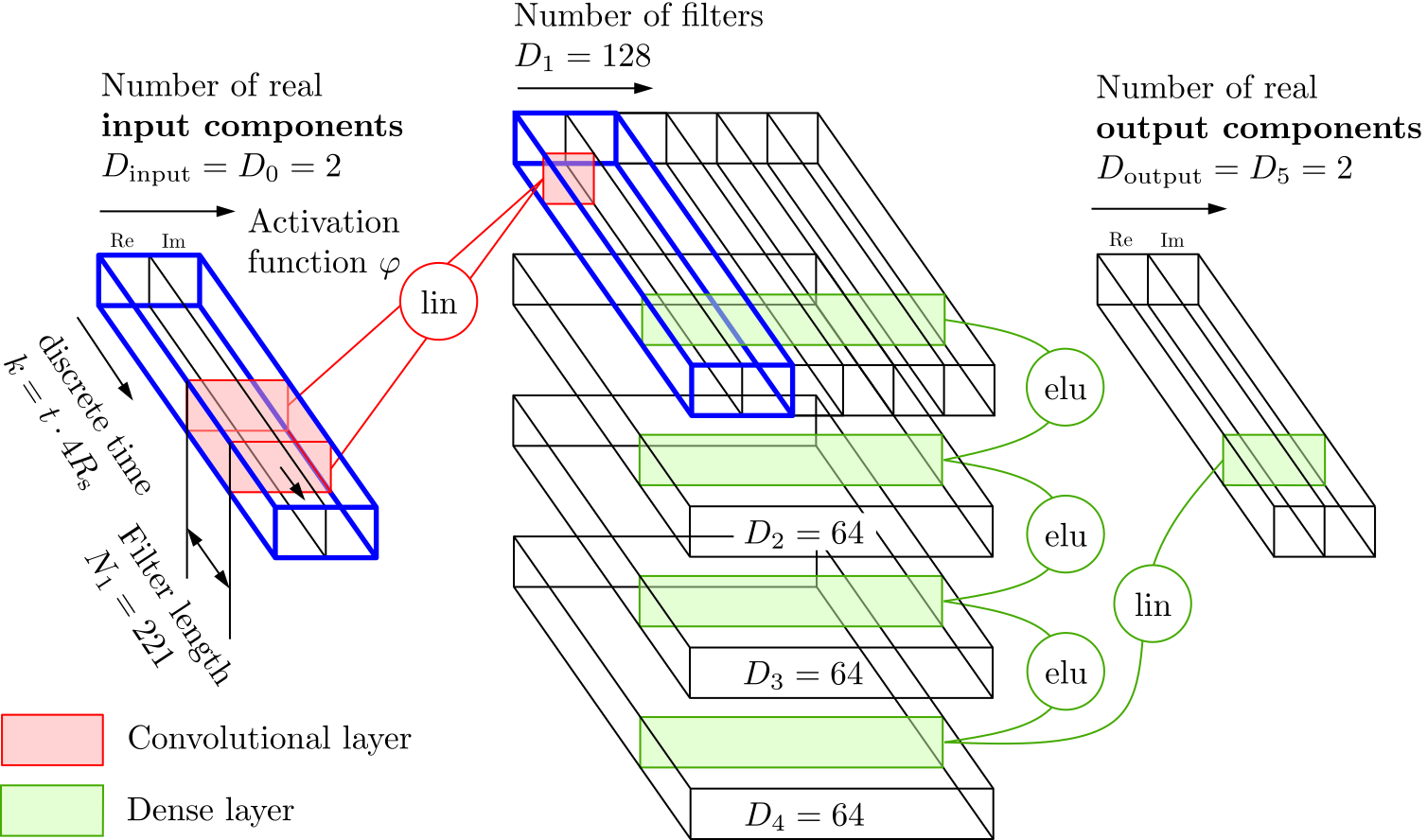}
	\caption{\acs{CNN} used at \ac{TX}. Boxes depict the signals $\mathbf{U}_\vartheta$. Bold blue lines depict the equivalent (linear) structure of \cite{uhlemann_deep-learning_2020} with $D_\mathrm{output}=D_1=2$.}
	\label{fig:cnn}
\end{figure}

The design of the \ac{TX} \ac{CNN} of Fig.~\ref{fig:aecnn-tx} is shown in Fig.~\ref{fig:cnn}, with the parameters given in Tab.~\ref{tab:cnn-configuration}. It is motivated by the fact that neighboring symbols nonlinearily interact with each other along the fiber. 
The input signal of the \ac{CNN} is $\mathbf{U}_0=\mathbf{C}_\mathrm{Re}\in \mathbb{R}^{u_\mathrm{sam}N_\mathrm{B}\times 2}$. Hence, the input length $K_0=u_\mathrm{sam}N_\mathrm{B}$ is the sequence of constellation symbols upsampled by $u_\mathrm{sam}$. The input dimension $D_0=2$ represents real and imaginary part of the original complex signal. 
The first layer is a convolutional layer that convolves its $D_1=128$ different filter kernels $\mathbf{H}_{1,d}\in\mathbb{R}^{N_1\times D_0}$ for $d<D_1$ with the real and imaginary part of the input signal. The length
\begin{equation}
	N_1=\left\lceil \Delta t_\mathrm{CD}\cdot R_\mathrm{s}\right\rceil \cdot u_\mathrm{sam}+1=221
\end{equation}
is chosen such that the filter length is in the order of the pulse widening induced by \ac{CD} of
\begin{equation}
	\Delta t_\mathrm{CD}=2\pi \beta_2\ell B_\mathrm{w}.
\end{equation}
Also the dimension $D_1$ is derived from the pulse widening such that this layer is \emph{at least} able to shift all relevant neighboring symbols to the same time instance by a (probably learned) time shifting filter $\mathbf{H}_{\mathrm{shift}}=\left(\delta_{-D_1,k},\,\ldots,\,\delta_{D_1,k}\right)$ with Kronecker-Delta
\begin{equation}
\delta_{n,k}=\left\{
\begin{array}{ll}
	1 & k = n \\
	0 & \text{else} \\
\end{array}
\right.
\end{equation}
and $k\in \left\{0,\,\ldots,\,K_1-1\right\}$. Hence, we have chosen $D_1=128\gg\left\lceil \Delta t_\mathrm{CD}\cdot R_\mathrm{s} \right\rceil=55$.
It follows a sub-network consisting of dense layers that operate on the single time instances of the $D_1$ convolved signals. The last layer again produces $D_5=D_0=2$ time signals that form the real and imaginary part of the pre-distorted signal $\mathbf{x}_\mathrm{CNN}=\mathbb{R}^2\text{2}\mathbb{C}\left\{\mathbf{U}_5 \right\}$. 

Bold blue lines mark the \ac{TX}-structure of the former \ac{AE} applied in \cite{uhlemann_deep-learning_2020}. Hence, the case $D_1=2$ and an imposed constraint on the kernel to perform a complex convolution, corresponds to the earlier \ac{TX}, i.e., the former structure is a special case of the one provided here, respectively.

\ifdefined\journal
\begin{figure*}[t] 
	\centering
	\includesvg[width=0.9\textwidth]{aect-rx-nn}
	\caption{Neural network of the \ac{RX} of Option 1 and the former \ac{AE} from \cite{uhlemann_deep-learning_2020}.}
	\label{fig:aect-rx-nn}
\end{figure*}
\fi

\subsection{Receiver Design}

The computational efficiency of the overlap \& save mechanism from \cite{uhlemann_deep-learning_2020} was not sufficient to be exhaustively trained such that we have had to deviate more from this intuitive design to reduce complexity. Hence, what follows is a \ac{CNN} that substitutes the overlap \& save mechanism in a more efficient way while having at least the same degree of freedom. Nevertheless, the achieved gains of the new architecture stem from the \ac{TX}. 

The \ac{RX} as shown in Fig.~\ref{fig:aecnn-rx-ii} consists of blocks for downsampling the receive signal $\mathbf{y}$ from simulation frequency $f_\mathrm{sim}$ to symbol rate $R_\mathrm{s}$, a complex-to-real-transformation, and a \ac{CNN} to learn an estimate of the desired likelihoods $\widehat{l}_{k,m}=\widehat{P}(S_k=m|\mathbf{y})$ of the transmitted symbol indices $S_k$ based on the received signal $\mathbf{y}$. Note that $S_k$ denotes the \ac{RV} corresponding to the true transmitted symbol index $s_k$ at time instance $k$ and a realization $m\in\mathcal{M}$. This can be obtained by a row-wise softmax-normalization of the \ac{CNN}'s output $[\mathbf{L}]_{k,m}=\widehat{l}_{k,m}$ and appropriate cost function as described below. The whole block can be written as a trainable function such that $\widehat{l}_{k,m}=g_\mathrm{RX}(k,m,\mathbf{y};\Omega_\mathrm{RX})$ with parameters $\Omega_\mathrm{RX}$ that consists of the weights $\mathbf{H}_\vartheta$ of the \ac{RX}-\ac{NN}. In a last step, the symbol with highest likelihood is chosen to be the hard decided estimate
\begin{equation}
	\widehat{s}_k=\arg\max_{m}g_\mathrm{RX}(k,m,\mathbf{y};\Omega_\mathrm{RX})\in\mathcal{M}.
	\label{eq:rx-estimate}
\end{equation}

The \ac{CNN} has the same architecture as in Fig.~\ref{fig:aecnn} but with different parameters, given in Tab.~\ref{tab:cnn-configuration}. The number of output neurons of the last layer is $M$ and has a linear activation function. The softmax-normalized output, then, can be interpreted as the previously introduced likelihoods of the transmit symbol.

\mode<all>
\mode*

\mode<article>{
\section{Training using \texorpdfstring{$\gamma$}{gamma}-Lifting}\label{sec:gamma-lifting}}

\mode<presentation>{
\section{Training using $\gamma$-Lifting}\label{sec:gamma-lifting}}

\begin{frame}
	\frametitle{Only by the aid of $\gamma$-lifting a stable convergence is obtained.}
	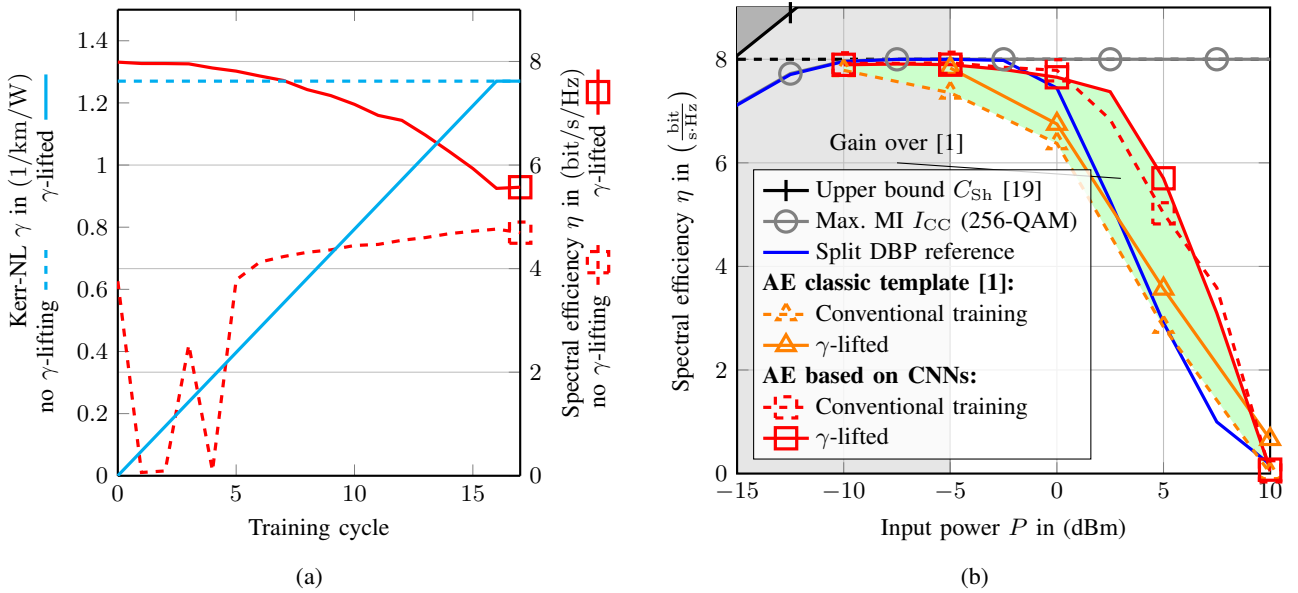
\begin{figure*}
		\only<2->{
			\begin{subfigure}[t]{0.5\textwidth}
			\newcommand{\eachnthpoint}{1}
			\centering
			\tikzset{external/export next=false}
			
			\pgfplotstableset{col sep=semicolon}
			\pgfplotstablevertcat{\output}{\results 20211214080021+20211215074837/training_20211214080021.csv}
			\pgfplotstablevertcat{\output}{\results 20211214080021+20211215074837/training_20211215074837.csv}			
			
			\begin{tikzpicture}[scale=1.0, spy using outlines={rectangle, black, magnification=3, size=1cm, connect spies}]
				\pgfplotsset{
					xmin=0, xmax=17,
					width=\columnwidth/1.25,
					height=22em,
					axis line style = thick,
					set layers,
					mark layer=axis background
				}
				
				\begin{axis}[title={},
					font=\small,
					xlabel={Training cycle},
					axis y line* = right,
					ymajorgrids = true,
					ylabel near ticks,
					xlabel near ticks,
					ymin=0,   ymax=9,
					ylabel style = {align = right},
					ylabel = {Spectral efficiency $\eta$ in $\left({\mathrm{bit}}/{\mathrm{s}}/{\mathrm{Hz}}\right)$\\ no $\gamma$-lifting \pgfplotsplotfromname{pgfplots:se-gamma-not-lifted} \hspace{1em} $\gamma$-lifted \pgfplotsplotfromname{pgfplots:se-gamma-lifted}},
					]				
					\only<3->{
					\addplot[very thick, color=red, mark=square, mark size=4pt, mark repeat=1, mark phase=18] table [col sep=semicolon, x expr=\coordindex, y expr=(\thisrowno{331}), each nth point=\eachnthpoint, filter discard warning=false, unbounded coords=discard] {\output}; 
					\label{pgfplots:se-gamma-lifted}}
					
					\addplot[dashed, very thick, color=red, mark=square, mark size=4pt, mark repeat=1, mark phase=18] table [col sep=semicolon, x expr=\coordindex, y expr=(\thisrowno{331}), each nth point=\eachnthpoint, filter discard warning=false, unbounded coords=discard] {\training{20211215145949}}; 
					\label{pgfplots:se-gamma-not-lifted}
				\end{axis}
				
				\begin{axis}[
					title={}, 
					font=\small,
					axis y line*=left,
					ylabel near ticks,
					axis x line=none,
					ymin=0,   ymax=1.5,
					ylabel style = {align = right},
					ylabel={Kerr-NL $\gamma$ in $\left(1/\mathrm{km}/\mathrm{W}\right)$\\ no $\gamma$-lifting \pgfplotsplotfromname{pgfplots:gamma-not-lifted} \hspace{1em} $\gamma$-lifted \pgfplotsplotfromname{pgfplots:gamma-lifted}},
					]
					
					\only<3->{
					\addplot[very thick, color=cyan] table [col sep=semicolon, x expr=\coordindex, y expr=(\thisrowno{20}), each nth point=\eachnthpoint, filter discard warning=false, unbounded coords=discard] {\output}; 
					\label{pgfplots:gamma-lifted}}
					
					\addplot[dashed, very thick, color=cyan] table [col sep=semicolon, x expr=\coordindex, y expr=(\thisrowno{20}), each nth point=\eachnthpoint, filter discard warning=false, unbounded coords=discard] {\training{20211215145949}}; 
					\label{pgfplots:gamma-not-lifted}
				\end{axis}
			\end{tikzpicture}
			\mode<article>{\subcaption{}}
			\label{fig:loss-over-iteration}
		\end{subfigure}}%
		~
		\begin{subfigure}[t]{0.5\textwidth} 
			\centering
			\tikzset{external/export next=false}
			\begin{tikzpicture}[scale=1.0, spy using outlines={rectangle, black, magnification=2, size=1cm, connect spies}]
				\begin{axis}[width=1.0\columnwidth,
					height=22em,
					font=\small,
					legend cell align={left},
					legend style={font=\small, fill opacity=0.8, text opacity=1.0},
					legend columns=1,
					legend pos=south west,
					axis line style = thick,
					xlabel={Input power $P$ in (dBm)},
					ylabel={Spectral efficiency $\eta$ in $\left(\frac{\mathrm{bit}}{\mathrm{s\cdot Hz}}\right)$},
					grid=major,
					xlabel near ticks,
					ylabel near ticks,
					xmin=-15.0,   xmax=10,
					ymin=0,   ymax=9,
					nodes near coords={
						\tiny
					},]
					
					\begin{pgfonlayer}{background}
						\draw[fill=black!10] (axis cs:-15, 0) rectangle (axis cs:-5, 9);
					\end{pgfonlayer}
					
					\addplot [very thick, forget plot, black, mark=none, dashed] coordinates {(-30, 8) (10, 8)};
					
					\mode<article>{\addplot[very thick, color=black, mark=|, mark size=4pt, mark repeat=1, mark phase=0, name path=SHANNON] table [x expr=10*log10(\thisrowno{33}*1000), y index = 326, col sep=semicolon] {\results 20211119160424/evaluation_20211119160424.csv};
					\addlegendentry{{Upper bound $C_\mathrm{Sh}$ \cite{kramer_upper_2015}}};}
					
					\mode<article>{\addplot[draw=none, forget plot, mark=none, dashed, name path=INF] coordinates {(-30, 10) (10, 10)};
					\addplot[forget plot, black!30] fill between [of=INF and SHANNON];
					
					\addplot[very thick, color=gray, mark=o, mark size=4pt, mark repeat=2, mark phase=0] table [x expr=10*log10(\thisrowno{33}*1000), y index = 328, col sep=semicolon] {\results 20211119160424/evaluation_20211119160424.csv};
					\addlegendentry{{Max. MI $I_\mathrm{CC}$ (256-QAM)}};}
					
					\addplot[very thick, color=blue, mark=none, mark size=4pt, mark repeat=2, mark phase=0] table [x expr=10*log10(\thisrowno{33}*1000), y index = 328, col sep=semicolon] {\results 20211119152924/evaluation_20211119152924.csv};
					\addlegendentry{Split DBP reference};			
					
					\addlegendimage{empty legend}
					\addlegendentry{\hspace{-.7cm}\textbf{AE classic template \cite{uhlemann_deep-learning_2020}:}};
					\addplot[name path=OLD, dashed, color=orange, mark=triangle, mark size=4pt, mark repeat=1, mark phase=0, very thick, restrict x to domain=-10:10] table [x index = 0, y index = 1] {\aecdnltwentyse};
					\addlegendentry{Conventional training};
					
					\only<4->{
					\addplot[very thick, color=orange, mark=triangle, mark size=4pt, mark repeat=1, mark phase=0] table [meta index=3, x expr=10*log10(\thisrowno{33}*1000), y index = 312] {\results 20211013105616/training_20211013105616.csv};
					\addlegendentry{$\gamma$-lifted};}
					
					\addlegendimage{empty legend}
					\addlegendentry{\hspace{-.7cm}\textbf{AE based on CNNs:}};
					
					\addplot[dashed, very thick, color=red, mark=square, mark size=4pt, mark repeat=2, mark phase=0] table [col sep=semicolon, meta index=3, x expr=10*log10(\thisrowno{33}*1000), y index = 331] {\results 20211129101104/training_20211129101104.csv};
					\addlegendentry{Conventional training};
					
					\only<4->{
					\addplot[name path=NEW, very thick, color=red, mark=square, mark size=4pt, mark repeat=2, mark phase=0] table [col sep=semicolon, meta index=3, x expr=10*log10(\thisrowno{33}*1000), y index = 331] {\results 20211206133958/training_20211206133958.csv};
					\addlegendentry{$\gamma$-lifted};
					
					\addplot[green!20] fill between[of=NEW and OLD];
					\node[above] at (axis cs:-7.5, 6) {\small{Gain over \cite{uhlemann_deep-learning_2020}}};
					\draw (axis cs:-7.5, 6) -- (axis cs:3, 5.7);}
				\end{axis}
			\end{tikzpicture}
			\mode<article>{\subcaption{}}
			\label{fig:results-se}
		\end{subfigure}
		\mode<article>{\caption{Performance evaluation of the \acs{AE} based on \acs{CNN} with (\subref{fig:loss-over-iteration}) showing the $\gamma$-lifting that describes the increase of $\gamma$ over training cycle on the left axis and the resulting \acs{SE} on the right axis at $P=5\,\mathrm{dBm}$; and (\subref{fig:results-se}) over input power compared to the different \acsp{AE} with $\gamma$-lifting (solid) and without (dashed).}}
	\end{figure*}
\end{frame}

\ifdefined\journal
\begin{figure}[t] 
	\centering
	\tikzset{external/export next=false}
	\begin{tikzpicture}[scale=1.0, spy using outlines={rectangle, black, magnification=2, size=1cm, connect spies}]
		\begin{axis}[width=1.0\columnwidth,
			height=22em,
			font=\small,
			legend cell align={left},
			legend style={font=\small, fill opacity=0.8, text opacity=1.0, at={(0.5,-0.2)}, anchor=north},
			legend columns=1,
			axis line style = thick,
			xlabel={Input power $P$ in (dBm)},
			ylabel={Spectral efficiency $\eta$ in $\left(\frac{\mathrm{bit}}{\mathrm{s\cdot Hz}}\right)$},
			grid=major,
			xlabel near ticks,
			ylabel near ticks,
			xmin=-30.0,   xmax=10,
			ymin=0,   ymax=9,
			nodes near coords={
			},]
			
			\begin{pgfonlayer}{background}
				\draw[fill=black!10] (axis cs:-30, 0) rectangle (axis cs:-10, 9);
			\end{pgfonlayer}
			
			\addplot [very thick, forget plot, black, mark=none, dashed] coordinates {(-30, 8) (10, 8)};
			\node[above] at (axis cs:-24, 8) {\small{$M_\mathrm{b}=8\,\mathrm{\frac{bit}{symbol}}$}};
			
			\addplot[very thick, color=black, mark=|, mark size=4pt, mark repeat=1, mark phase=0, name path=SHANNON] table [x expr=10*log10(\thisrowno{33}*1000), y index = 326, col sep=semicolon] {\results 20211119160424/evaluation_20211119160424.csv};
			\addlegendentry{{Upper bound $C_\mathrm{Sh}$ \cite{kramer_upper_2015}}};
			
			\addplot[draw=none, forget plot, mark=none, dashed, name path=INF] coordinates {(-30, 9) (10, 9)};
			\addplot[forget plot, black!30] fill between [of=INF and SHANNON];
			\node[above, align=center] at (axis cs:-24, 6.3) {Unattainable\\region};
			
			\addplot[very thick, color=gray, mark=o, mark size=4pt, mark repeat=2, mark phase=0] table [x expr=10*log10(\thisrowno{33}*1000), y index = 328, col sep=semicolon] {\results 20211119160424/evaluation_20211119160424.csv};
			\addlegendentry{{Max. MI $I_\mathrm{CC}$ (256-QAM)}};
			
			\addplot[very thick, color=blue, mark=none, mark size=4pt, mark repeat=2, mark phase=0] table [x expr=10*log10(\thisrowno{33}*1000), y index = 328, col sep=semicolon] {\results 20211119152924/evaluation_20211119152924.csv};
			\addlegendentry{Split DBP reference};
			
			\addplot[very thick, color=green, mark=none, mark size=4pt, mark repeat=1, mark phase=0, very thick] table [x expr=10*log10(\thisrowno{30}*1000), y index = 89] {\classiccdnlse};
			\addlegendentry{Static channel equalization \cite{savory_digital_2010}};
			
			\addlegendimage{empty legend}
			\addlegendentry{\hspace{-.7cm}\textbf{AE classic template \cite{uhlemann_deep-learning_2020}:}};
			
			\addplot[dotted, color=orange, mark=triangle, mark size=4pt, mark repeat=1, mark phase=0, very thick, restrict x to domain=-10:10] table [x index = 0, y index = 1] {\aecdnltwentyse};
			\addlegendentry{conventional training};

			\addlegendimage{empty legend}
			\addlegendentry{\hspace{-.7cm}\textbf{AE based on CNNs:}};
			
			\addplot[dotted, very thick, color=red, mark=square, mark size=4pt, mark repeat=2, mark phase=0] table [col sep=semicolon, meta index=3, x expr=10*log10(\thisrowno{33}*1000), y index = 331] {\results 20211126085236/training_20211126085236.csv};
			\addlegendentry{conventional training};
			
			\addplot[dashed, very thick, color=red, mark=square, mark size=4pt, mark repeat=2, mark phase=0] table [col sep=semicolon, meta index=3, x expr=10*log10(\thisrowno{33}*1000), y index = 331] {\results 20211129101104/training_20211129101104.csv};
			\addlegendentry{reset training};
			
		\end{axis}
	\end{tikzpicture}
	\caption{Performance of the new \ac{AE} (red) using a conventional and reset training compared with the former linear architectural template of \cite{uhlemann_deep-learning_2020} (orange).}
	\label{fig:results-se-conventional}
\end{figure}
\fi

The general objective of the trainable \ac{AE} is an optimal estimate of the transmitted symbols at \ac{RX}
\begin{equation}
	\widehat{\mathbf{s}}\stackrel{!}{=}{\mathbf{s}} \Rightarrow \widehat{s}_k\stackrel{!}{=}s_k\,\forall\,k,
\end{equation} 
where $\mathbf{s}$ is the vector containing the transmitted symbol indices, and $\widehat{\mathbf{s}}$ the corresponding estimate by the \ac{RX}. 

To achieve this goal, the parameters $\Omega=\left\{\Omega_\mathrm{TX},\,\Omega_\mathrm{RX}\right\}$ of \ac{TX} and \ac{RX} are trainable, which is common for all \ac{AE}-setups in communications. Hence, not only \ac{TX} and \ac{RX}, but the whole \ac{AE} system of Fig.~\ref{fig:system} can be seen as a trainable function $g(k,m,\mathbf{s};\Omega)$. This way, the estimate of the symbol at time instance $k$ is the argument of the maximization of the learned probabilities over all symbol indices $\widehat{s}_k=\arg\max_{m}g(k,m,\mathbf{s};\Omega)$ with $m\in\mathcal{M}$.
Having said this, the objective can be further detailed to $g(k,m,\mathbf{s};\Omega)=\widehat{P}(S_k=m)\stackrel{!}{=} P(S_k=m)\,\forall\,k$, where
\begin{equation}
	P(S_k=m)=\begin{cases}
		1, &m=s_k \\
		0, &\text{else}
	\end{cases}
	\label{eq:true-distribution}
\end{equation}
is the \ac{PMF} of the true transmit symbol indices $s_k=\left[\mathbf{s}\right]_k$ (this refers to ``one-hot encoding'', as the \ac{PMF} of the transmitted symbols corresponds to the Kronecker-delta). The optical channel is contained in $g(k,m,\mathbf{s};\Omega)$ and serves as the penalty to overcome by choosing $\Omega$ appropriately.

Equivalent to \cite{uhlemann_deep-learning_2020}, we trained the \ac{AE} using samplewise \ac{CE} loss for batch sample $k$ defined as
\begin{equation}
	\begin{gathered}
		\mathcal{L}_\mathrm{CE}\left({P}(S_k=m),\,\widehat{P}(S_k=m)\right) \\
		\begin{aligned}
			&=\mathrm{H}_{P(S_k=m)}\left[\widehat{P}(S_k=m)\right] \\
		\end{aligned}
	\end{gathered}
\end{equation}
where $\mathrm{H}[\cdot]$ is \acl{CE} and $\mathrm{E}[\cdot]$ is expectation. Taking into account Eq.~(\ref{eq:true-distribution}) this can be further simplified to
\begin{equation}
	\begin{gathered}
		\mathcal{L}_\mathrm{CE}\left({P},\,\widehat{P}\right) = -\log g(k,m=s_k,\mathbf{s};\Omega)
	\end{gathered}
\end{equation}
The resulting cost is defined over the batch of symbols with size $N_B$ and calculates as
\begin{equation}
	\begin{gathered}
		\mathcal{{C}}_\mathrm{CE}\left({P}(S_k=m),\,\widehat{P}(S_k=m)\right) \\
		\begin{aligned}
			&=-\frac{1}{N_\mathrm{B}}\sum^{N_\mathrm{B}-1}_{k=0}\log g(k,m=s_k,\mathbf{s};\Omega)
		\end{aligned}
	\end{gathered}
\end{equation}
For the optimizer we have chosen Adam \cite{kingma_adam_2017}, which applies \ac{SGD} to minimize the cost as
\begin{equation}
	\begin{gathered}
		\Omega_\mathrm{opt}=\arg\min_{\Omega}\mathcal{{C}}_\mathrm{CE}\left({P},\,\widehat{P}\right)
	\end{gathered}
\end{equation}
by training the parameters $\Omega$. 
For a conventional training as in \cite{uhlemann_deep-learning_2020} the channel is fixed and only training parameters, i.e., the learning rate $u_\mathrm{lr}$ and batch size $N_\mathrm{B}$, are adjusted during training to support convergence. The latter is achieved by the aid of the training configuration of Tab.~\ref{tab:conventional-training}.

\begin{table}[b] 
	\caption{Configuration of a single training cycle.}
	\small
	\begin{center}
		\begin{tabular}{c|c|c}
			\textbf{Max. training iterations}&\textbf{Batch size}&\textbf{Learning rate} \\
			$N_\mathrm{train}$&$N_\mathrm{B}$&$u_\mathrm{lr}$ \\
			\hline
			2000&12,000&$2\cdot10^{-4}$ \\
			1500&12,000&$1\cdot10^{-4}$ \\
			1000&12,000&$5\cdot10^{-5}$ \\
			500&12,000&$1\cdot10^{-5}$
		\end{tabular}
		\label{tab:conventional-training}
	\end{center}
\end{table}

\ifdefined\journal
In Fig.~\ref{fig:results-se-conventional} the results of this conventional training are shown. The \ac{AE} based on a classical template from \cite{uhlemann_deep-learning_2020} was retrained with the configuration of Tab.~\ref{tab:conventional-training} (dotted orange) and compared to the \ac{SCE} from \cite{savory_digital_2010} (green) as before. We have already shown that the \ac{AE} can outperform the \ac{SCE} while exhibiting a comparable complexity.

In this work, we have now included the \ac{DBP} as a reference system (blue), which achieves a \ac{SE} $\eta_\mathrm{DBP}\approx 7\,\mathrm{bit/s/Hz}$ at $P=0\,\mathrm{dBm}$ that is much higher than $\eta_\mathrm{SCE}\approx2\,\mathrm{bit/s/Hz}$ of the \ac{SCE} and even higher than the retrained \ac{AE} based on a \ac{CT} from \cite{uhlemann_deep-learning_2020} $\eta_\mathrm{AECT}\approx 6\,\mathrm{bit/s/Hz}$. This was expected as the used split \ac{DBP} consists of a nonlinear \ac{PD}, whereas the \ac{TX} of the \ac{AE} based on a \ac{CT} can only linearly transform the bare transmit signal. In contrast, also the trained \ac{AE} with \acp{CNN} (dotted red) exhibits no gain compared to the former \ac{AE}, which may only be due to an insufficient training procedure. Simple adjustments of the training parameters as, e.g., given in Tab.~\ref{tab:conventional-training} did not further yield reliable improvements.

The light gray area indicates that we are not interested in this region, where the nonlinearity is not significant.

As the performance of the new \ac{AE} seems to vary over input power one may conclude that it did not converge well. Hence, we tried to stabilize the convergence, or, at least increase the probability of convergence by repeating the training described by Tab.~\ref{tab:conventional-training} and hence, resetting the learning rate periodically. The set of training iterations between two resets is called a cycle.
Unfortunately, this was not (exhaustively) possible with \ac{RX} Option 1, as its inefficient overlap \& save mechanism requires too much time. Hence, we switch (from now on) to \ac{RX} Option 2. Fig.~\ref{fig:loss-over-iteration} shows the \ac{SE} over all training cycles in dashed violet for \ac{RX} Option 1 and dashed red for \ac{RX} Option 2. Obviously, as intended, no gain was introduced by \ac{RX} Option 2 (we only expect a gain introduced by the nonlinear \ac{TX}). Further, indeed, the convergence is not always stable but happens after several cycles. This ``stabilized convergence'' can also be seen in dashed red in  Fig.~\ref{fig:results-se-conventional}. This way, we are able to outperform the split \ac{DBP} in blue at high input powers $P\geq \SI{0}{\dBm}$. There may be other conventional training strategies that may improve the probability of convergence. Nevertheless, we will make use of a fundamentally different approach.
\fi

Nevertheless, the very first results have shown that the training procedure of \cite{uhlemann_deep-learning_2020} is not sufficient to train the proposed architectural template. Only by repeating the training given in Tab.~\ref{tab:conventional-training}, by chance, we obtain the dashed red results of Fig.~\ref{fig:results-se}. The underlying problem is depicted also in dashed red in Fig.~\ref{fig:loss-over-iteration}, where one can see that the performance jumps during the first training cycles, i.e., repetitions.
Having realized that convergence is a major problem of training the \ac{AE} through the \ac{SSFM}, we further investigated on how to increase the stability of the convergence and to probably find better local optima. Hence, in the following we introduce a new training method that was inspired by the training described in \cite{tandler_recurrent_2019}. There, the input training samples were chosen such that the learning gets simplified or, i.e., the relation between input and output becomes more obvious to the trainable structure.

We have searched for a way to simplify the stated problem such that we can gradually converge until finally arriving at the desired more complex problem to solve. Here, this may be achieved by setting some channel parameters to zero and increasing them incrementally, which, in effect, results in an \emph{increasing difficulty} for training the system. Thereby, we ensure that training converges in each cycle, before making it harder. As potential parameters we identified the following:

Slowly increasing fiber length $\ell$ simplifies the problem significantly as all impairments increase with fiber length. On the downside, it poses an unsolvable conflict between a varying accuracy and a vanishing gradient information. This can be seen by considering the step size of the \ac{SSFM} $\Delta z_\mathrm{SSFM} = {\ell}/{N_\mathrm{SSFM}}$. A growing $\ell$ means either an increasing $\Delta z_\mathrm{SSFM}$ and hence a varying accuracy; or adaption of $N_\mathrm{SSFM}$, which means that the error has to be backpropagated through an increasing number of \ac{SSFM}-iterations. This way, the very first training cycles may be predominant compared to the later ones with weaker gradient information.
Using chromatic dispersion $\beta_2$ means reducing \ac{ISI} at the beginning and incrementally adding more influence of neighboring symbols. This allows to start with a simple sample-wise compensation at the beginning, but is in contrast to the requirement that the filter length $N_\vartheta$ of the \ac{CNN} has to be fixed to the final length over all trainings, anyway. Hence, for the first training cycles, most of the randomly initialized trainable filter taps rather distort than may ever support.
For considering Kerr-effect only, $\gamma$ controls the power-dependent rotation of a sample, where, in turn, its power depends on the corresponding transmit symbol, neighboring transmit symbols, and noise. Reducing $\gamma$ disentangles the simple linear superposition of \ac{CD} and may help to learn their interactions step-by-step.

As there is no obvious drawback connected to the Kerr-nonlinearity parameter, we decided to use a $\gamma$-lifting. This $\gamma$-over-cycle dependency is also depicted in Fig.~\ref{fig:loss-over-iteration} for $P=5\,\mathrm{dBm}$ by the solid cyan curve, which linearly increases with training cycle. Note that, each training cycle consists of all training iterations described by Tab.~\ref{tab:conventional-training}. The conventional training is depicted by the dashed cyan curve showing that $\gamma$ is constant. It may be misleading that the \ac{SE} worsens over training cycle, which is a consequence of the increasing difficulty. Nevertheless, comparing both red curves -- the one without $\gamma$-lifting (dashed) and the one with (solid) -- shows that the latter exhibits a significantly stabilized convergence, as no large variations can be observed anymore. Further, it achieved a significantly higher \ac{SE} for the final $\gamma=1.27\,\mathrm{km^{-1}W^{-1}}$, i.e., $5.7\,\mathrm{bit/s/Hz}$ vs. $5.0\,\mathrm{bit/s/Hz}$.

Performing $\gamma$-lifting over the whole input power space results in Fig.~\ref{fig:results-se}. The dashed curves again show the earlier \ac{AE} (orange) and the one proposed in this work (red) with their conventional training. Switching on $\gamma$-lifting leads to the corresponding solid curves.
It can be seen that $\gamma$-lifting, in most cases, pushes the performance of the \ac{CNN}-based \ac{AE} closer to $I_\mathrm{CC}$. By chance, thereby, the conventional training may still exceed the $\gamma$-lifting due to the high fluctuations in the training quality. Nevertheless, the proposed \ac{AE} outperforms the former \ac{AE} as well as the split \ac{DBP} at high input powers. This means that we have compensated or avoided spectral broadening partly and may operate at higher launch powers, which may be beneficial for higher modulation formats or reach. Even by the application of $\gamma$-lifting to this former \ac{AE}, its performance stays well below the one proposed here, due to the limitations of its linear architectural template. The \ac{CNN}-based \ac{AE} is not able to reach the numerical limit that was introduced by Fig.~\ref{fig:dbp-reference}, but significantly contributes to closing this gap.

\mode<all>
\mode*
\section{Summary and conclusion}\label{sec:summary}

\mode<presentation>{
	
	\begin{frame}
		\frametitle{Outlook: Lumped instead of Raman amplification plus WDM.}
		\begin{figure}[ht]
			\centering
			\includesvg[width=0.8\textwidth]{lumped-amplification}
			\label{fig:lumped-amplification-channel}
		\end{figure}	
		\begin{figure}
			\centering
			\includegraphics[width=0.4\textwidth]{fig/png/classicdbptrxwdm-wdm}
			\label{fig:dscm-tx-signal-presentation}
		\end{figure}
	\end{frame}
	
	\begin{frame}
		\frametitle{Outlook: First results are promising.}
		\begin{figure}[h] 
			\centering
			\begin{tikzpicture}[scale=1.0, spy using outlines={rectangle, black, magnification=1.5, size=1.2cm, connect spies}]
				\begin{axis}[%
					se chart,
					height=22em,
					width=0.7\textwidth,
					xlabel={Signal-to-noise-ratio $\mathrm{SNR}$ in $\left(\mathrm{dB}\right)$},
					xmin=10,
					xmax=40,
					ymin=0,
					axis x line*=top,
					grid=none,
					]
				\end{axis}
				\begin{axis}[%
					se chart,
					height=22em,
					width=0.7\textwidth,
					xlabel={Input power per subcarrier $P_\mathrm{sub}$ in (dBm)},
					ylabel={Spectral efficiency $\eta$ in $\left(\frac{\mathrm{bit}}{\mathrm{s\cdot Hz}}\right)$},
					xmin=-20,
					ymin=0,
					domain=-20:10,
					]
					
					\begin{pgfonlayer}{background}
					\end{pgfonlayer}
					
					\addplot [thick, forget plot, black, mark=none, dashed] coordinates {(-30, 8) (10, 8)};
					\node[above] at (axis cs:-15, 8) {\small{$M_\mathrm{b}=\SI{8}{\bit\per\second\hertz}$}};
					
					\addplot[forget plot, curve, color=black, mark=|, name path=SHANNON] {log2(1+10^((x+30.156)/10))};
					
					\addplot[draw=none, forget plot, mark=none, dashed, name path=INF] coordinates {(-30, 9) (10, 9)};
					\addplot[forget plot, black!30] fill between [of=INF and SHANNON];
					
					\addplot[forget plot, curve, color=black, dashed, mark=none, name path=REF] table [x expr=10*log10(\thisrowno{51}/\thisrowno{100}*1000), y index = 2, col sep=semicolon] {\evaluation{20220429131913}{20220429131913}};
					\addplot[forget plot, curve, color=black, dotted, mark=none, name path=LPF] table [x expr=10*log10(\thisrowno{51}/\thisrowno{100}*1000), y index = 2, col sep=semicolon] {\evaluation{20220429122318}{20220429122318}};
					
					\addlegendentry{Chrom.-Disp.-only};
					\addplot[curve, color=green, mark=x] table [x expr=10*log10(\thisrowno{53}/\thisrowno{102}*1000), y index = 2, col sep=semicolon] {\training{20220502151655}};
					
					\addlegendentry{Per subcarrier};
					\addplot[curve, color=red, mark=x] table [x expr=10*log10(\thisrowno{53}/\thisrowno{102}*1000), y index = 2, col sep=semicolon] {\training{20220502204721}};
					
					\addlegendentry{With neighboring info.};
					\addplot[curve, color=blue, mark=x] table [x expr=10*log10(\thisrowno{53}/\thisrowno{102}*1000), y index = 2, col sep=semicolon] {\training{20220505072758}};
					
					\addplot[forget plot, yellow!30] fill between [of=REF and LPF, soft clip={domain=-10:10}];
					\node at (axis cs:5, 6) {\blockbox{3cm}{Potential gain\\for the AE}};
					
				\end{axis}
			\end{tikzpicture}
		\end{figure}
	\end{frame}

\begin{frame}
\frametitle{The learnings of the Autoencoder are feasible and comparable to DBP.}
\begin{itemize}
	\item We have learned a nonlinear pulse shaping for the Autoencoder's transmitter.
	\item Training the systems requires $\gamma$-lifting for a good convergence.
	\item The learned system can outperform DBP where it is limited by spectral broadening.
	\item We are now ready to apply the Autoencoder to full fledged DWDM system.
\end{itemize}
\end{frame}
}

In this work, we have identified spectral broadening introduced by Kerr-nonlinearity as one of the major performance limitation of a \ac{DBP} for the simulated optical fiber channel with bandlimited \ac{TX} and \ac{RX}.

The nowadays well-known \ac{AE} was chosen as an approach to learn to address this challenge and to find an appropriate nonlinear pulseshaping. We showed that the \ac{TX}-part of the \ac{AE} can be trained through the \ac{SSFM}. As it turned out, a stable convergence of the trainable structures requires a carefully adjusted training method. We have therefore combined a conventional training with a novel ``$\gamma$-lifting'' strategy to achieve significant gains compared to the currently best known \ac{NLC} schema, i.e., split \ac{DBP}, in a regime where spectral broadening is significant.

For now, we have only considered a single channel and single polarization scenario, to obtain a better understanding of the effects of Kerr-nonlinearity. In future work, these ideas can be extended to a multi-channel system (\ac{WDM}), or by applying even more sophisticated architectural templates.

\ifdefined\journal
\begin{figure}[t] 
	\centering
	\tikzset{external/export next=false}
	\begin{tikzpicture}[scale=1.0, spy using outlines={rectangle, black, magnification=1.5, size=1.2cm, connect spies}]
		\begin{axis}[title={}, 
			width=1.0\columnwidth,
			height=30em,
			font=\small,
			legend cell align={left},
			legend style={font=\small, fill opacity=0.9, text opacity=1.0,},
			legend columns=1,
			legend pos=south west,
			axis line style = thick,
			xlabel={Input power $P$ in (dBm)},
			ylabel={Spectral efficiency $\eta$ in $\left(\frac{\mathrm{bit}}{\mathrm{s\cdot Hz}}\right)$},
			grid=major,
			xlabel near ticks,
			ylabel near ticks,
			xmin=-30.0,   xmax=10,
			ymin=0,   ymax=14,
			nodes near coords={
			},]
			
			\begin{pgfonlayer}{background}
				\draw[fill=black!10] (axis cs:-30, 0) rectangle (axis cs:-10, 14);
			\end{pgfonlayer}
			
			\addplot [very thick, forget plot, black, mark=none, dashed] coordinates {(-30, 8) (10, 8)};
			\node[above] at (axis cs:-24, 8) {\small{$M_\mathrm{b}=8\,\mathrm{\frac{bit}{symbol}}$}};
			
			\addplot[very thick, color=black, mark=|, mark size=4pt, mark repeat=1, mark phase=0, name path=SHANNON] table [x expr=10*log10(\thisrowno{33}*1000), y index = 326, col sep=semicolon] {\evaluation{20211119160424}{20211119160424}};
			\addlegendentry{{Upper bound $C_\mathrm{Sh}$ \cite{kramer_upper_2015}}};
			
			\addplot[draw=none, forget plot, mark=none, dashed, name path=INF] coordinates {(-30, 14) (10, 14)};
			\addplot[forget plot, black!30] fill between [of=INF and SHANNON];
			\node[above, align=center] at (axis cs:-24, 6.3) {Unattainable\\region};
			
			\addplot[very thick, color=gray, mark=o, mark size=4pt, mark repeat=2, mark phase=0] table [x expr=10*log10(\thisrowno{33}*1000), y index = 328, col sep=semicolon] {\evaluation{20211119160424}{20211119160424}};
			\addlegendentry{{Max. MI $I_\mathrm{CC}$ (256-QAM)}};
			
			\addplot[forget plot, draw=none, name path=A, no markers, mark options={decoration={name=none}}, decoration={text along path,
				text={{without LPF}},raise=+8pt,
				text align={left indent={0.76\dimexpr\pgfdecoratedpathlength\relax}},
			},
			postaction={decorate}] table [x expr=10*log10(\thisrowno{33}*1000), y index = 328, col sep=semicolon] {\evaluation{20211119160424}{20211119160424}};
			
			\addlegendimage{empty legend}
			\addlegendentry{\hspace{-.7cm}\textbf{Split DBP {\color{red}without LPF}:}};
			
			\addplot[very thick, color=green!40!black, mark=none, mark size=4pt, mark repeat=1, mark phase=0] table [x expr=10*log10(\thisrowno{33}*1000), y index = 327, col sep=semicolon] {\evaluation{20211217114913}{20211217114913}};			
			\addlegendentry{{$N_\mathrm{SSFM}=600$}};	
			
			\addplot[very thick, color=green!60!black, mark=none, mark size=4pt, mark repeat=1, mark phase=0] table [x expr=10*log10(\thisrowno{33}*1000), y index = 327, col sep=semicolon] {\evaluation{20211217103444}{20211217103444}};
			\addlegendentry{{$N_\mathrm{SSFM}=400$}};
			
			\addplot[very thick, color=green!80!black, mark=none, mark size=4pt, mark repeat=1, mark phase=0, name path=HIGH] table [x expr=10*log10(\thisrowno{33}*1000), y index = 327, col sep=semicolon] {\evaluation{20211217095804}{20211217095804}};
			\addlegendentry{{$N_\mathrm{SSFM}=200$}};
			
			
			\addlegendimage{empty legend}
			\addlegendentry{\hspace{-.7cm}\textbf{Split DBP with LPF:}};
			
			\addplot[very thick, color=blue, mark=none, mark size=4pt, mark repeat=2, mark phase=0, name path=LOW] table [x expr=10*log10(\thisrowno{33}*1000), y index = 328, col sep=semicolon] {\evaluation{20211119152924}{20211119152924}};
			\addlegendentry{{$N_\mathrm{SSFM}=200$}};
			
			\addplot[very thick, color=violet, mark=none, mark size=4pt, mark repeat=2, mark phase=0] table [x expr=10*log10(\thisrowno{36}*1000), y index = 443, col sep=semicolon] {\evaluation{20220325224524}{20220325224524}};
			\addlegendentry{{$M=1024$}};
			
			\addplot[very thick, color=pink, mark=none, mark size=4pt, mark repeat=2, mark phase=0] table [x expr=10*log10(\thisrowno{36}*1000), y index = 442, col sep=semicolon] {\evaluation{20220328071604}{20220328071604}};
			\addlegendentry{{$M=1024$ no LPF}};
			
			\addplot[forget plot, yellow!30] fill between [of=HIGH and LOW];
			
			\addplot[forget plot, draw=none, name path=A, no markers, mark options={decoration={name=none}}, decoration={text along path,
				text={{with LPF}},raise=+4pt,
				text align={left indent={0.7\dimexpr\pgfdecoratedpathlength\relax}},
			},
			postaction={decorate}] table [x expr=10*log10(\thisrowno{33}*1000), y index = 328, col sep=semicolon] {\evaluation{20211119152924}{20211119152924}};
		\end{axis}
	\end{tikzpicture}
	\caption{Comparison of the performance of a split \ac{DBP} with and without (DAC/ADC-induced) \acp{LPF}. The numerical accuracy is sufficient such that it is the \acp{LPF} that limit performance.}
\end{figure}
\fi

\ifCLASSOPTIONcaptionsoff
  \newpage
\fi



\bibliographystyle{IEEEtranTCOM}
\bibliography{IEEEabrv,usr/references}
%
%
%
%

%




\end{document}